\pdfoutput=1
\RequirePackage{fix-cm}
\documentclass[final]{svjour3}                    

\usepackage[top=2cm, bottom=2cm, left=2cm, right=2cm]{geometry}
\usepackage[numbers]{natbib}

\usepackage[T1]{fontenc}

\usepackage{booktabs} 
\usepackage{xcolor}
\usepackage{amsmath}

\usepackage{bm} 

\usepackage{amsfonts}
\usepackage{balance}
\usepackage{multirow}
\usepackage{graphicx}
\usepackage{caption}
\usepackage{subcaption}
\usepackage{url}
\usepackage{dsbda}
\usepackage{setspace}

\usepackage[utf8]{inputenc}
\usepackage[english]{babel}

\usepackage{blindtext}

\definecolor{temp}{HTML}{FF0000}
\definecolor{todo}{HTML}{FFA419}
\definecolor{done}{HTML}{0BF9AB}
\definecolor{info}{HTML}{1E90FF} 
\definecolor{moresamples}{HTML}{00FF00}

\renewcommand{\train}{\text{train}}
\renewcommand{\test}{\text{test}}
\newcommand{\pred}{\text{pred}}

\newcommand{\mlp}{\operatorname{MLP-2}}
\newcommand{\Nexperiments}{4,590}

\begin{document}

\title{Recommendations for Item Set Completion: On the Semantics of Item~Co-Occurrence With Data Sparsity, Input Size, and Input Modalities}
\titlerunning{Recommendations for Item Set Completion}
            
\author{I. Vagliano        \and
        L. Galke 
        \and A. Scherp 
}


\institute{I. Vagliano \at
              Amsterdam University Medical Centers, Amsterdam, the Netherlands \\
              \email{i.vagliano@amsterdamumc.nl}           
           \and
           L. Galke \at
              Kiel University and ZBW - Leibniz Information Centre for Economy, Kiel, Germany \\
              \email{lga@informatik.uni-kiel.de} 
          \and
          A. Scherp, University of Ulm, Ulm, Germany \\
          \email{ansgar.scherp@uni-ulm.de} 
}

\date{}

\maketitle

\begin{abstract}
We address the problem of recommending relevant items to a user in order to ``complete'' a partial set of items already known.
We consider the two scenarios of citation and subject label recommendation, which resemble different semantics of item co-occurrence: relatedness for co-citations and diversity for subject labels.
We assess the influence of the completeness of an already known partial item set on the recommender performance.
We also investigate data sparsity through a pruning parameter, and the influence of using additional metadata.
As recommender models, we focus on different autoencoders, which are particularly suited for reconstructing missing items in a set. 
We extend autoencoders to exploit a multi-modal input of text and structured data. 
Our experiments on six real-world datasets show that supplying the partial item set as input is helpful when item co-occurrence resembles relatedness, while metadata are effective when co-occurrence implies diversity.
This outcome means that the semantics of item co-occurrence is an important factor.
The simple item co-occurrence model is a strong baseline for citation recommendation.
However, autoencoders have the advantage to enable exploiting additional metadata besides the partial item set as input, and achieve comparable performance.
For the subject label recommendation task, the title is the most important attribute.
Adding more input modalities sometimes even harms the result.
In conclusion, it is crucial to consider the semantics of the item co-occurrence for the choice of an appropriate recommendation model and carefully decide which metadata to exploit.

\end{abstract}

\section{Introduction}

We address the problem of recommending items to a user in order to ``complete'' a partial set of items that the user already knows.
Such a general recommendation task finds its applications in research paper recommendation~\cite{beel2016paper, DBLP:journals/ipm/RaamkumarFP17}, citation recommendation~\cite{caragea2013can,DBLP:conf/sigir/EbesuF17,zhao_citation_2021}, and multi-label classification~\cite{DBLP:journals/jdwm/TsoumakasK07}, in the scientific digital library community known as subject indexing~\cite{iso999}.
For the latter, the task is for professional indexers to choose representative annotations from a domain-specific thesaurus in order to label scientific papers.
The existing body of works is typically concerned with one-off recommendation, \eg recommendations of scientific collaborators~\cite{DBLP:journals/ir/ZhouDLW17}, suggestion of which paper or news to read next~\cite{beel2016paper, DBLP:journals/ipm/HuLSYS20, DBLP:journals/ir/CucchiarelliMSV19}, or recommending an open sequence of items, \eg music recommendations on Spotify, which are continuously extended as long as the users are listening~\cite{10.1145/2652481,DBLP:conf/recsys/VaglianoGMS18, DBLP:journals/ir/WangDX18}.

In contrast, we consider recommendation scenarios where there are items to be recommended to complete an existing, partial set of items.
This  set of items is complete at some point in time.
For example, when writing a scientific article, at some point in time the set of cited papers is accomplished for the newly authored paper.
This notion of completeness does not guarantee to have every possible related article covered with the citations, \ie some papers may be missing due to various reasons, but the authored paper itself is finished and with it, the set of citations is concluded.
In these set completion tasks, the number of possible outputs is as large as the power set of possible items, such that emitting crisp decisions, as usual in multi-label classification, is notoriously challenging~\cite{DBLP:conf/www/TangRN09,DBLP:conf/nips/NamMKF17,DBLP:conf/icml/NamKMPSF19}. When we frame these problems as recommendation tasks, we rank the items rather than take a hard decision, and authors or subject indexers may benefit from recommendations until they consider the set to be complete.

\begin{figure}[!h]
  \begin{subfigure}[b]{0.5\textwidth}
       \centering
       \includegraphics[width=0.8\textwidth]{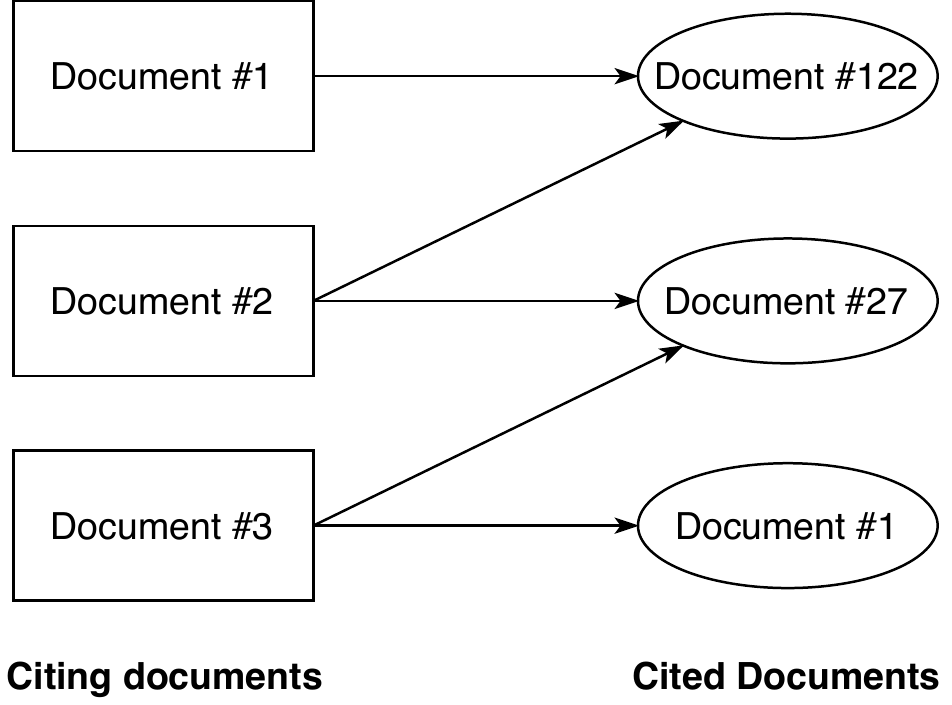}
        \caption{The citation recommendation scenario.}
    \end{subfigure}
    \quad 
    \begin{subfigure}[b]{0.5\textwidth}
       \centering
        \includegraphics[width=0.8\textwidth]{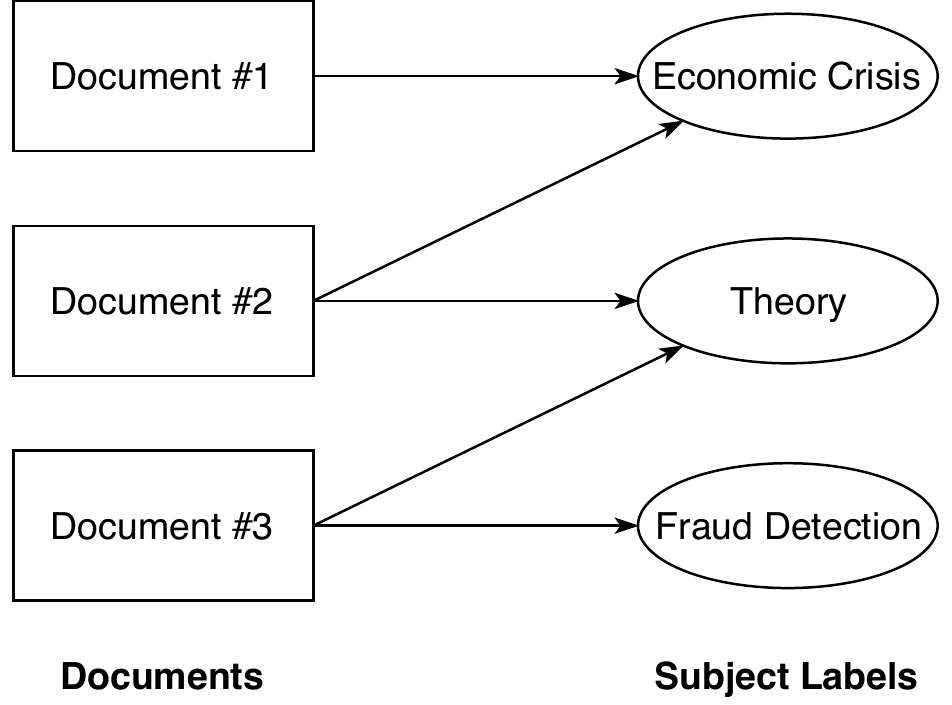}
        \caption{The subject indexing scenario.} 
    \end{subfigure}
  \caption{Exemplary bipartite graphs of citation relationships between documents
    (left) and documents annotated with subject labels (right)~\cite{Galke:2018}.
  The two recommendation tasks share a similar structure, but in the former the recommendation items are the cited documents, in the latter are the subject labels.}\label{fig:example-viz}
\end{figure}

We consider two recommendation scenarios (Figure~\ref{fig:example-viz}): Citation recommendation and subject label recommendation.
They have been chosen as they have different, complementing semantics of items occurring together in a set (co-occurring items).
While in citation recommendation item co-occurrence implies similarity of the cited papers, in professional subject indexing, co-occurrence means dissimilarity of the annotated subjects.
Thus, for citation recommendation, similar items will be sought, while labels that are different from those already assigned are naturally good recommendations for subject indexing. Despite the different underlying semantics, both scenarios can be modeled in a common framework.
We take either the citations or the assigned labels as implicit feedback for the considered recommendation task.

\subsection{Research Questions}

In our experiments, we are interested in understanding the performance of recommending items to be added to an existing partial set under different conditions: 

\textit{(i)~Is there a difference in the recommender performance, when there are different underlying semantics of item co-occurrence in a set?}
We chose citation and subject label recommendation as our scenarios because of the difference in what it means when two items occur together.

\textit{(ii)~How does the completeness of the partial set of items influences the provided recommendations? 
Is it easier for the recommender to suggest good items at an early, mid, or late completeness stage for the set of items?} 

We analyze the impact of the size of the partial set given as input by varying the number of dropped elements that need to be predicted for the output.
The more items are dropped, the fewer items are available as input to the recommender (but more  choices are available as correct items for the recommender).
This question naturally renders the challenge of recommending items to a user to complete an existing, partial set of items, given different levels of completeness of this set, which represent different stages in the task of finding citations or subject labels.

\textit{(iii)~How does pruning of rarely-cited documents, as well as documents which cite few other documents, affect the models' performance?}
Existing studies are often applied on datasets where rarely cited documents and documents that cite too few other works are removed~\cite{beel2016paper}. 
This prevents verifying how the recommender will behave in real settings.
To systematically investigate how the pruning threshold affects the models' performance, we conduct experiments where the pruning threshold is a controlled variable.

\textit{(iv)~
What is the optimal use of additional data known about the documents?}
We investigate the influence of different metadata attributes such as authors, title, venue on the recommendation performance, while using the publication year to create a chronological train-test split.

\subsection{Models and Evaluation Results}

As recommender models, we focus on autoencoder architectures and compare them to strong baselines.
Autoencoders have become very popular in recent time for recommendation tasks~\cite{zhao_citation_2021,DBLP:journals/www/PanHY20,DBLP:conf/www/LiangKHJ18,DBLP:conf/www/Steck19}, sometimes using side information~\cite{DBLP:conf/icmlc2/BaiB19,DBLP:journals/access/HeMZ19,DBLP:conf/recsys/ChenR18,DBLP:journals/bigdatama/LiuWKH18}.
By learning to reconstruct their input, autoencoders are useful to recommend items for the set completion task.
Due to regularization, autoencoders can capture general patterns in the data rather than merely copying their input~\cite{bengio2013representation}.
For our experiments,
we consider four representative techniques to regularize autoencoders, namely undercomplete, denoising, variational, and adversarial autoencoders.

We extend the autoencoders to take multi-modal input such that they become  hybrid recommenders that can exploit both the ratings in terms of the partial item set as well as the content in terms of additional metadata.
These extended models can receive different metadata fields as input, such as the documents' title, authors, and venue. 
We compare autoencoders with three strong baselines for a fair evaluation and to address the question of whether neural architectures are making progress regarding the considered recommendation tasks~\cite{DBLP:conf/recsys/DacremaCJ19}.
As baselines, we have selected a multi-layer perceptron (MLP), a classical recommendation model based on singular value decomposition (SVD), and a simple yet strong item co-occurrence baseline.
We chose MLP because it is known from prior works to perform well in subject label classification tasks~\cite{Galke:2018,DBLP:conf/jcdl/MaiGS18}.

We evaluate our models and baselines in both recommendation scenarios on three different datasets for each scenario.
In total, we use six different datasets, which stem from five domains, namely medicine, computer science, economics, politics, and news.
We pre-process the documents by splitting them into a training and testing set along the time axis, at a point $t$, \ie our models are being trained on documents published before $t$ and tested on the documents published after.  
This resembles the natural constraint that newly written publications can only cite already published works.
We apply the same choronological split to subject labels to account for concept drift~\cite{DBLP:journals/datamine/WebbLGP18}, \ie changes in the joint distribution of documents and annotations over time.
These settings are challenging as they correspond to a cold-start situation, more specifically, the well-known new user problem~\cite{DBLP:journals/is/SilvaCPMR19}.
We investigate each scenario regarding the influence of different proportions of completeness of the partial item set for the performance and different pruning of documents.
Furthermore, we experiment with different combinations of metadata attributes as side information to the recommender.
We run $\Nexperiments{}$ experiments.

The key takeaway from our experiments is that supplying the partial item sets as input is only helpful, when item co-occurrence resembles relatedness, while the content is effective when co-occurrence implies diversity.
Thus, when facing a new recommendation task, it is crucial to consider the semantics of item co-occurrence for the choice of an appropriate model and metadata attributes to exploit.
Our results support this claim: In citation recommendation (co-occurring items are similar), the performance of the strong item co-occurrence baseline can only be matched by using additional metadata. In subject indexing (co-occurring items are diverse), using the content, \eg with MLP or conditioned autoencoders, is much more effective for predicting missing items than merely using the partial item set. We have confirmed these results on different degrees of sparsity (via controlled dataset pruning) and at different stages of the iterative set completion process.

\subsection{Contributions and Structure of the Article}

Below, we summarize our contributions.
It is a significantly extended research based on own prior work~\cite{Galke:2018}.
We added the research questions on the influence of the size of the partial input set, we used more datasets and models, which we also extend to exploit different metadata field in a multi-modal fashion, and we provide a deeper analysis of how metadata affects the recommendation tasks.

\begin{enumerate}
  \item We define a common representation for recommendation tasks in the context of set completion, where item co-occurrence resembles different semantics of the relation between items.
  This representation is based on the two real-world scenarios of citation recommendation, where item co-occurrence means relatedness, and subject label recommendation, where co-occurrence implies diversity. 
  
  \item We investigate the performance of the recommendation models along four factors:
  (i)~the semantics of item co-occurrence, (ii)~the influence of the completeness of the partial set of items used as input, (iii)~the pruning of infrequent items, and (iv)~the optimal use of additional metadata available about the items.
  We extend different recommender architectures to incorporate metadata in addition to a partial set of items as input in a multi-modal fashion.
  
  \item As experimental models, we focus on autoencoder architectures.
  They resemble a natural choice for the set completion task and are very popular for recommendation tasks.
  We compare four autoencoder architectures with strong neural as well as non-neural baselines, to answer the four factors (i) to (iv).
  This contributes to the pressing question whether neural architectures are making a contribution to the state of the art in recommender systems~\cite{DBLP:conf/recsys/DacremaCJ19}.
  
  \item  We evaluate all autoencoder architectures and baselines in the two scenarios on six datasets from five different domains.
  The experimental data is split along the time axis, to resemble real-world settings.
  
\end{enumerate}

In Section~\ref{sec:rw}, we review previous work on recommender systems with a focus on autoencoder-based approaches and methods for citation and subject recommendation. 
In Section~\ref{sec:problem-statement}, we formally state the recommendation problem and introduce the extended multi-modal autoencoder models and baselines.
We describe the experimental apparatus for the citation and subject recommendation experiments in Section~\ref{sec:experiments}. 
We present the results our experiments in Section~\ref{sub:results} and discuss the results in Section~\ref{sec:discussion}, before we conclude.

\section{Related Work}\label{sec:rw}

We review previous work on recommender systems with a focus on autoencoder-based approaches.
Then, we discuss methods and systems specifically designed for citation and subject label recommendation. 

\subsection{Relation to Collaborative Filtering, Content-based, and Hybrid Recommender Systems}

Recommender systems that recommend items to a user taking into account ratings
that users with similar preferences gave to these items, i.e. operate only on the rating matrix $\sU \times \sI$, are typically labeled as \emph{collaborative filtering}~\cite{Felfernig2013}.
On the other hand, approaches that take item content or ratings that users gave to items into account are referred to as \emph{content-based}~\cite{Lops2011}.
\emph{Hybrid} techniques combine these two approaches and consider both the content and user-item ratings.
In hybrid recommenders, one can further distinguish between loose coupling and tight coupling~\cite{DBLP:conf/kdd/WangWY15}.
In loose coupling, collaborative filtering and content-based models are separate and their final output is combined, whereas in tight coupling, a joint model operates on both input modalities.
Our approaches behave as collaborative-filtering recommenders when only the partial set is given, while are content-based when only the metadata is given, and they behave as tightly coupled hybrid approaches when both the partial set and additional metadata are used.

\subsection{Autoencoders for Recommendation Tasks}

Autoencoders have recently gained wide popularity for recommendation tasks~\cite{DBLP:journals/www/PanHY20,DBLP:conf/www/LiangKHJ18,DBLP:conf/www/Steck19,DBLP:conf/ACISicis/CaoYL17,DBLP:journals/nn/ZhuangZQSXH17,DBLP:conf/www/SedhainMSX15}, oftentimes using side information~\cite{DBLP:conf/icmlc2/BaiB19,DBLP:journals/access/HeMZ19,DBLP:conf/recsys/ChenR18,DBLP:journals/bigdatama/LiuWKH18,DBLP:conf/kdd/WangWY15,DBLP:conf/ijcnn/MajumdarJ17,DBLP:conf/kdd/LiS17,DBLP:conf/iconip/ZhangYXWZ17,DBLP:conf/recsys/StrubGM16,DBLP:conf/cikm/LiKF15}.
Autoencoders are especially well-suited for tight-coupling collaborative filtering and content-based recommendations~\cite{DBLP:conf/icmlc2/BaiB19,DBLP:conf/kdd/WangWY15,DBLP:conf/kdd/LiS17}.
A common strategy is to combine auto-encoded item features with a latent factor model of the ratings~\cite{DBLP:conf/icmlc2/BaiB19,DBLP:journals/access/HeMZ19,DBLP:conf/kdd/LiS17}. 
Some works also fuse user/item side-information with the respective rows/columns of the rating matrix as input to the autoencoder~\cite{DBLP:conf/recsys/ChenR18,DBLP:conf/ijcnn/MajumdarJ17}.
Majumdar and Jain~\cite{DBLP:conf/ijcnn/MajumdarJ17} investigated the pure cold start problem and partial cold start problems, in which they assume that 10\% or 20\% of the ratings are present.
In contrast, we investigate additional stages of rating completeness.
Regarding architectures, the variational autoencoder (VAE)~\cite{KingmaW13} is recently most-often used for recommendation tasks~\cite{DBLP:conf/www/LiangKHJ18,DBLP:conf/icmlc2/BaiB19,DBLP:journals/access/HeMZ19,DBLP:conf/recsys/ChenR18,DBLP:conf/kdd/LiS17}, although other works have used (stacked) denoising autoencoders~\cite{DBLP:journals/bigdatama/LiuWKH18,DBLP:conf/kdd/WangWY15,DBLP:conf/cikm/LiKF15}.
Autoencoders find widespread applications in recommender systems.
The most prominent variants~\cite{DBLP:conf/www/LiangKHJ18,DBLP:conf/kdd/LiS17} stood the test of a recent study on reproducibility in deep learning recommender systems~\cite{DBLP:conf/recsys/DacremaCJ19}, that has shown that all tested autoencoder variants, namely Collaborative VAE~\cite{DBLP:conf/kdd/LiS17} and Mult-VAE~\cite{DBLP:conf/www/LiangKHJ18} could be reproduced.
 
Compared to previous works, we focus on using autoencoders for the new user problem~\cite{DBLP:journals/is/SilvaCPMR19} as modeled by a chronological split of the datasets along the time axis. 
In contrast, most existing literature assumes that all users are already present during training.
Furthermore, we explicitly investigate the influence of different semantics of item co-occurrence on the recommendation performance.
We also consider different completeness levels of the partial set in our experiment, beginning from early to mid to late stages of the two scenarios of citation and subject label recommendation.

\subsection{Research paper and citation recommendation}\label{sec:citrec}
Research paper recommendation is a well-known and popular topic~\cite{beel2016paper,ali_overview_2021}. 
Specifically for citation recommendation~\cite{2020arXiv200206961F}, one distinguishes between recommendations
based on a partial set of references and recommendations based on the content
of a manuscript~\cite{DBLP:conf/cikm/HuangKCMGR12}. 
While the former strives to identify missing citations on the broader document level, the latter is suited for finding matching citations for a given statement during writing.
Citation recommendation recently focuses on these context-sensitive applications,
in which concrete sentences are mapped to, preferably relevant,
citations~\cite{DBLP:journals/scientometrics/MaZL20,DBLP:conf/sigir/EbesuF17,zhao_citation_2021,beel2016paper,DBLP:conf/cikm/HuangKCMGR12}.
Instead, we revisit the reference set completion problem and we do not take the context of the citation into account, as the full text of a paper is rarely available in large-scale metadata sources~\cite{DBLP:conf/jcdl/MaiGS18}. 
Co-citation analysis assumes that two papers are the more related to each other, the
more they are co-cited~\cite{DBLP:journals/jasis/Small73}.
Following that idea, Caragea et al. relied on singular value
decomposition as a more efficient and extendable approach for citation recommendation~\cite{caragea2013can}.
Other approaches make use of deep learning techniques for citation recommendation but focus on context-sensitive scenarios~\cite{DBLP:conf/sigir/EbesuF17,Huang:2015:NPM:2886521.2886655,DBLP:journals/access/SakibAH20,DBLP:journals/access/ZhangM20b, DBLP:journals/cin/TaoSZD20,DBLP:journals/el/ChenYLC20,DBLP:journals/eswa/AliKMAI20}. 
Thus, we recognize the need for new methods that are not only based on item co-occurrence but also take additional metadata into account for these partial set completion problems. 
While in our own preliminary study~\cite{Galke:2018}, we have considered only title data, we now investigate the benefit of using more metadata, investigate the influence of different sizes of partial item sets, pruning of the items, and generalize the results on further datasets and models.

Further approaches from the areas of network analysis include node embeddings~\cite{DBLP:conf/kdd/PerozziAS14,node2vec-kdd2016}, link prediction with graph neural networks~\cite{DBLP:conf/nips/ZhangC18},  and dynamic graph representation learning~\cite{JODIE}. However, these methods and also retrieval methods such as IRGAN~\cite{10.1145/3077136.3080786} do not apply to our problem because they require all documents to be known in advance, while our methods need to be applicable to new unseen data without further training. In graph representation learning, the former is known as transductive learning, while the latter is called inductive learning~\cite{HamiltonBook}.

\subsection{Subject Label Recommendation}

Subject label recommendation is similar to tag recommendation: in both cases
the goal is to suggest a descriptive label for some content.
Boughareb et al. show a new approach to recommend tags for scientific papers, which defines the relatedness between the tags attributed by users and the concepts extracted from the available sections of scientific papers based on statistical, structural and semantic aspects~\cite{DBLP:journals/ijcis/BougharebKBFS20}.
Sun et al. presented a hierarchical attention model for personalized tag recommendation~\cite{DBLP:journals/jasis/SunZJLW21}.
Lei et al. introduce a tag recommendation by text classification that uses the capsule network with dynamic routing for tag recommendation.~\cite{DBLP:journals/ijon/LeiFYL20} The capsule network encodes the intrinsic spatial relationship between a part and a whole constituting viewpoint invariant knowledge that automatically generalizes to novel viewpoints.
Zhou et al. propose a novel hybrid method based on multi-modal content analysis that recommends keywords for video upload to compose titles and tags~\cite{DBLP:journals/access/ZhouXWZ20}. They combine textual semantic analysis of original tags and recognition of video content with deep learning.
Similarly, Sigurbj\"ornsson et al. proposed a tag recommender for Flickr to support the user in the photo annotation
task~\cite{Sigurbjornsson:2008:FTR:1367497.1367542},
whereas Posch et al. predict hashtag categories
on Twitter~\cite{Posch:2013:MCU:2487788.2488008}.
While these works focus on tags for social media, we consider subjects from a standardized thesaurus for scientific documents.

Tag or label recommendation is also related to the problem of multi-label classification.
Nam et al.~\cite{DBLP:conf/nips/NamMKF17} propose an encoder-decoder architecture based on sequence models as a special case of the generic set2set method~\cite{DBLP:journals/corr/VinyalsBK15}. 
They extend this approach by an Expectation-Maximization model to supply context-dependent label permutations to the sequence mode~\cite{DBLP:conf/icml/NamKMPSF19}. 
%
In our prior work, we analyzed titles vs full-text on scientific and news corpora for multi-label classification~\cite{DBLP:conf/kcap/GalkeMSBS17}. 
We found that using only the title retains 90\% of the accuracy obtained by using the full-text.
Among many baselines, a wide MLP with a single hidden layer is preferable. 
Furthermore, we showed that using only the titles can even be better than using the full-text for multi-label classification tasks, when more training data is available~\cite{DBLP:conf/jcdl/MaiGS18}.
This motivates us to use additional bibliographic metadata in the two recommendation scenarios, rather than the full-text, which is often not available even in open access datasets~\cite{DBLP:conf/jcdl/MaiGS18}.

\subsection{Summary}
Autoencoders have proven effective in recommendation tasks and have been widely used. 
To the best of our knowledge, no previous work has yet studied in detail the item recommendation at different stages of the partial set, \ie with a varying number of items in input, and in the new user settings.
Furthermore, we consider the differences in the semantics of item co-occurrence in the two recommendation scenarios of citation and subject label as well as different degrees of sparsity and the influence of using metadata.

\section{Problem Formalization and Models}\label{sec:problem-statement}

We first describe the two scenarios considered. Then, we provide a formal problem statement for the considered
recommendation tasks. 
The documents can be considered users in a traditional recommendation scenario, while the items are either cited documents or subject labels~\cite{Galke:2018}, respectively.
Subsequently, we present the used autoencoders models. 
We focus on autoencoders because they are specifically designed to reconstruct ``corrupted'' input. 
\subsection{Recommendation Scenarios}

Below, we present the two recommendation scenarios of citation recommendation and subject label recommendation and their semantics of item co-occurrence.

\paragraph{Scenario 1: Item Co-occurrence as Similarity in Citation Recommendation}
When writing a new paper, it is essential that the authors reference other publications which are key in the respective field of study or relevant to the paper being written. 
Failing to do so can be rated negatively by reviewers in a peer-reviewing process. 
However, due to an increasing volume of scientific literature, even some key papers are sometimes overlooked. 
We address this challenge by studying the problem of recommending publications to consider as citation candidates.
Given that the authors have already selected some references for their scientific paper, we use these references as partial input set for recommendations of further documents that may be included as references. 
Initially, the set is very short since the completeness of the citation set depends on the stage of writing, while towards submission is almost complete.
If a document is at an early stage, where few other documents are cited, the chance to find a proper recommendation is high, but the information about which candidates to recommend is sparse as the set so far is rather small.
In the middle stage of the writing process, more references have been included, but some more need to be found.
In a late stage, most of the relevant documents have been added, making the set almost complete and the choices for further good recommendations small. 
We explicitly consider in our experiments the influence of completeness of the partial input set on the recommendation performance. 

\paragraph{Scenario 2: Item Co-occurrence as Dissimilarity in Subject Indexing}
Subject indexing is a common task in scientific libraries to make scientific documents accessible for search. 
New documents are manually annotated by qualified subject indexers with a set of subject labels, i.\,e., classes from a, typically domain-specific, thesaurus.
Fully-automated multi-label classification approaches to subject indexing are promising~\cite{DBLP:conf/nips/NamMKF17}, even when merely the metadata of the publications is used~\cite{DBLP:conf/kcap/GalkeMSBS17}. 
Professional subject indexers, however, typically use the result of these approaches only as recommendations, so that the human-level quality can still be guaranteed. 
Similarly to the citation task, the professional subject indexers start adding labels one by one to the publications.
Following internationally standardized guidelines~\cite{iso999}, the indexers ensure that the labels are covering well the often multiple scientific aspects, contributions, and methods of the papers.
To this end, the indexers scan the paper's title, section headings, and partially read their content.
Thus, properly indexing a scientific paper with subject labels is a difficult task that can be supported by a recommender system.
The recommender takes the partial set of already assigned subject labels as input and generates recommendations for new labels that refer to aspects of the paper that are not yet covered.
Following the same experimental approach as for the citation task, we explicitly evaluate the influence of the different levels of completeness of the already assigned subject labels on the recommendation performance.
Thus, we use different sizes of the partial input set of labels and measure the quality of the recommendations.

\subsection{Problem Formalization}

Given a set of \( m \) documents, \(\mathbb D\), and
a set of \( n \) items, \(\mathbb I\), the typical recommendation task is to model
the spanned space, \(\mathbb D \times \mathbb I\).
We model the ratings as a sparse matrix \( \boldsymbol{X} \in { \lbrace 0,1 \rbrace } ^ {m \times
n}\), in which \( X_{jk} \) indicates implicit feedback from document \( j \) to item \( k \).
To simulate a real-world scenario, we split
the documents \( \mathbb{D} \) into \(m_\train \) documents for training, \(\mathbb{D}_\train \), and \( m_\test \) documents for evaluation, \(\mathbb{D}_\test \),
such that \(\mathbb{D}_\train \cap \mathbb{D}_\test =
\emptyset \). 
More precisely, we conduct this split into training and test documents based on
the publication year. All documents that were published before a certain year
are used as training, and the remaining documents as test data.
This leads to an experimental setup that is close to a real-world application for citation recommendation and subject label recommendation.
More details are provided in Section~\ref{sub:procedure}.
All models are supplied with the users' ratings \(\boldsymbol{X}_\train = \mathbb{D}_\train \bowtie \boldsymbol{X}\) along
with the side information \( \boldsymbol{S}_\train =
\mathbb{D}_\train \bowtie \boldsymbol{S}\) for training.
As side information $\boldsymbol{S}$, we use the documents' various metadata fields such as title, authors, and venue. 
The test set \(\boldsymbol{X}_\test, \boldsymbol{S}_\test \) is obtained analogously.
A summary of our notation can be found in Table~\ref{tab:notation}.

For evaluation, we remove randomly selected items in \( \boldsymbol{X}_\test \) by setting a fraction of the
non-zero entries in each row to zero. We denote the hereby created test set by \( \boldsymbol{\tilde X}_\test \).
The model ought to predict values \(\boldsymbol{X}_\pred \in {\mathbb{R}}^{m_\test \times n} \), given the test set, \( \boldsymbol{\tilde X}_\test \),
along with the title information, \( \boldsymbol{S}_\test \).
Finally, we compare the predicted scores, \(\boldsymbol{X}_\pred \), with the
true ratings, \( \boldsymbol{X}_\test \), via ranking metrics.
The goal is that those items, that were omitted in 
\( \boldsymbol{\tilde X}_\test \) are highly ranked in \( \boldsymbol{X}_\pred \).

\begin{table}[!h]
\small
  \caption{Notation table}\label{tab:notation}
  \centering
\begin{tabular}{ll}
  \toprule
  Symbol & Description \\
  \midrule
  $\sD$ & Set of \(m\) documents \\
  \(\mathbb I\) & Set of \(n\) items \\
  \(\boldsymbol{X} \in {\lbrace 0,1 \rbrace}^{m \times n}\) & Sparse ratings matrix \\
  \(\boldsymbol{S} \in {\mathbb{R}}^{m \times d}\) & Side information from document
  metadata\\
  \(\boldsymbol{x}, \boldsymbol{s} \) & Row vectors of \( \boldsymbol{X} \) or \( \boldsymbol{S} \), respectively \\
  \(\lbrack \boldsymbol{x}; \boldsymbol{s} \rbrack \) & Concatenation of vectors \(\boldsymbol{x} \) and \(\boldsymbol{s} \) \\
  \(\bowtie \) & Natural join (on document identifiers) \\
  \( \boldsymbol{I} \) &  Identity matrix \\
  \bottomrule
\end{tabular}
\end{table}

In both scenarios, \ie citation recommendation and subject label recommendation, we regard documents and items as a bipartite graph, as depicted in Figure~\ref{fig:example-viz}.
Considering citations, this point of view may be counter-intuitive since a scientific document is typically both a citing paper and a cited paper. 
Still, the out-degree of typical citation datasets is so high that there are an order of magnitude more papers in the set of cited papers than in the set of citing papers of the bipartite graph.
For instance, the PubMed citation dataset has 224,092 documents that cite 2,896,764 distinct other documents.
Therefore, it is reasonable to distinguish the documents based on their role in the citation relationship, i.\,e., citing versus cited paper.

\subsection{Autoencoder Models for Itemset Reconstruction}\label{sec:models}
Autoencoders are particularly suited for our task as they aim to reconstruct corrupted input. Here corruption means missing items. Thus, we focus on autoencoder architectures and compare them with strong baselines (introduced in Section~\ref{sec:baselines}). Below, we introduce the multi-layer perceptron as a building block for the autoencoders, and we show how metadata can be incorporated in the former as well as in undercomplete, denoising, variational and adversarial autoencoders.


\begin{figure*}[!h]
    \centering
    \begin{subfigure}[b]{0.45\textwidth}
        \centering
       \includegraphics[width=\textwidth]{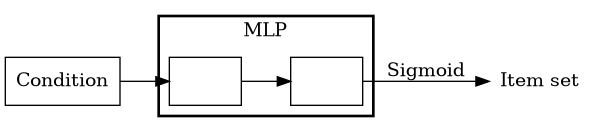}
        \caption{2-layer multi-layer perceptron (MLP)}
        \label{fig:mlp}
    \end{subfigure}
    
    \begin{subfigure}[b]{0.8\textwidth}
        \centering
       \includegraphics[width=\textwidth]{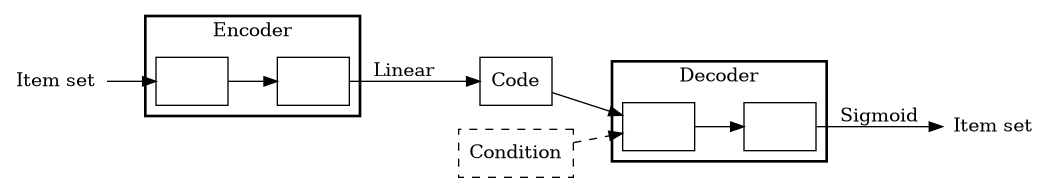}
        \caption{Undercomplete autoencoder}
        \label{fig:ae}
    \end{subfigure}
    
    \begin{subfigure}[b]{0.8\textwidth}
        \centering
       \includegraphics[width=\textwidth]{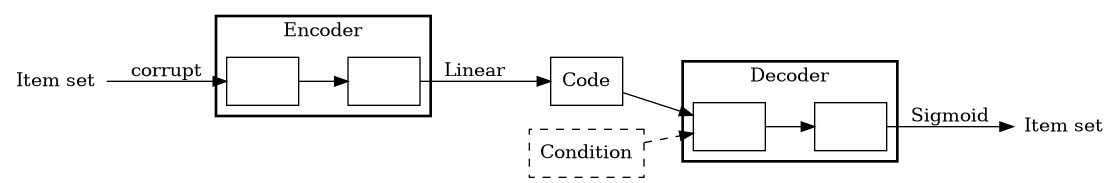}
        \caption{Denoising autoencoder}
        \label{fig:dae}
    \end{subfigure}
    
    \begin{subfigure}[b]{0.8\textwidth}
        \centering
       \includegraphics[width=\textwidth]{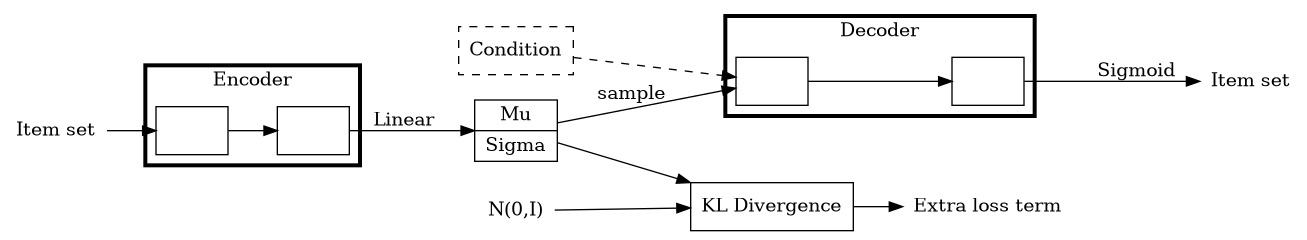}
        \caption{Variational autoencoder}
        \label{fig:vae}
    \end{subfigure}
    
    \begin{subfigure}[b]{0.8\textwidth}
        \centering
       \includegraphics[width=\textwidth]{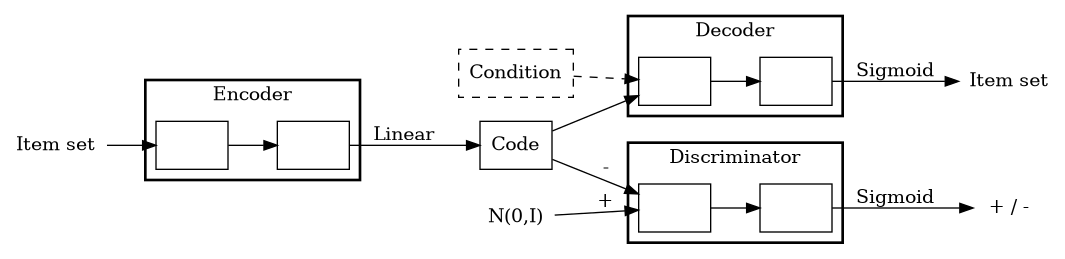}
        \caption{Adversarial autoencoder}
        \label{fig:aae}
    \end{subfigure}
  \caption{
  Considered autoencoder architectures based on a 2-layer MLP as building block during training:
  (a)~MLP baseline and its use as encoder and decoder in the autoencoders for item-based recommendations (b-e). 
  Each encoder/decoder/discrimination block is an MLP module.
  When not labeled differently, the activation function is ReLU followed by dropout. 
  The condition is the supplied metadata.
  Only denoising autoencoder (c) corrupts the input item set during training, while the other architectures train the code on the full item set.}
  \label{fig:aes}
\end{figure*}

\paragraph{Multi-Layer Perceptron}\label{sec:dist-rep-mlp}

A multi-layer perceptron (MLP) is a fully-connected feed-forward neural network
with one or multiple hidden layers (Figure~\ref{fig:mlp}). The output is computed by consecutive
applications of \(\vh^{(i)} = \sigma(\vh^{(i-1)} \cdot
\mW^{(i)} + \vb^{(i)})\), with \(\sigma\) being a nonlinear
activation function. 
In the description of the following models, we
abbreviate a two-hidden-layer perceptron module by MLP-2.

\paragraph{Undercomplete Autoencoders} An autoencoder (AE)
has two main components: the encoder \texttt{enc} and the decoder
\texttt{dec} (Figure~\ref{fig:ae}). 
The encoder transforms the input $\vx$ into a hidden
representation, the code
\( \vz = \operatorname{enc}(\vx) \). 
The decoder aims to reconstruct the input from the code \(\vr = \operatorname{dec}(\vz)\).
The two components are jointly trained to minimize 
the binary cross entropy, \(\operatorname{BCE}(\vx,\vr)\). To avoid learning to merely copy the input \(\vx\) to
the output \(\vr\), autoencoders need to be regularized. The most common
way to regularize autoencoders is by imposing a lower dimensionality on the code (undercomplete autoencoder). 
In short, autoencoders are 
trained to capture the most important explanatory factors of variation for
reconstruction~\cite{bengio2013representation}.
For both the encoder and the decoder we chose an MLP-2 module, such that the model function becomes \(
\vr = \operatorname{MLP-2_{dec}}(\operatorname{MLP-2_{enc}}(\vx)) \). 

\paragraph{Denoising Autoencoders} A denoising autoencoder (DAE) is an autoencoder that receives corrupted data as input and aims to predict original, uncorrupted data~\cite{Vincent:2008:ECR:1390156.1390294} (Figure~\ref{fig:dae}). During the training, the initial input $\bm{x}$ is corrupted into $\tilde{\vx}$ through stochastic mapping \( \tilde{\vx} \sim q_D(\tilde{\vx}|\vx) \), \eg randomly forcing a fraction of entries to zero (white noise).
The corrupted input \( \tilde{\vx} \) is then mapped to a hidden representation (the code) \( \vz = \operatorname{enc}(\tilde\vx) \) in the same way of the standard autoencoder, and from the hidden representation the model reconstructs \(\vr = \operatorname{dec}(\vz) \).
Both the encoder and the decoder rely on a MLP-2 module, such that the model function becomes \(
\vr = \operatorname{MLP-2_{dec}}(\operatorname{MLP-2_{enc}}(\tilde\vx)) \). 
The loss function is again binary cross-entropy. 

\paragraph{Variational Autoencoders} A variational autoencoder (VAE) is a generative model whose posterior is approximated by a neural network, forming an autoencoder architecture~\cite{KingmaW13,10.5555/3044805.3045035} (Figure~\ref{fig:vae}). Variational autoencoders use a variational approach for latent representation learning, in which the underlying data-generating distribution $p_\mathrm{data}(\rvx|\rvz)$ is assumed to be a mixture of latent variables $\vz$. The encoder \( q_\theta(\rvz|\rvx) \) learns to infer these latent variable, while the decoder \( p_\theta(\rvx|\rvz) \) learns to generate samples from these latent variables, where \( \phi \) and \( \theta \) are the parameters of the encoder and decoder, respectively. The goal is that $p_\theta(\vx|\vz)$ approximates $p_\text{data}$ after successful training. A crucial component of the VAE is the reparametrization trick that facilitates backpropagation through random operations~\cite{KingmaW13}. Given a selected prior distribution $p_\text{prior}(\rvz)$ on the code $\rvz$, a deterministic encoder learns to predict the parameters of this distribution. 
The prior distribution $p_\text{prior}(\rvz)$ of a VAE is typically Gaussian~\cite{KingmaW13}. The deterministic output of the encoder $\operatorname{enc}(\vx)$ is first split into two halfs: one for the mean, $\mu$, and one for the standard deviations, $\sigma$.
Then, we independently sample $\epsilon \sim \mathcal{N}(0,1)$ and compose the code as $\vz = \mu + \sigma \epsilon$ to be fed into the deterministic decoder.
To encourage a normal distribution on the code, the Kullback-Leibler (KL) divergence with respect to $\mathcal{N}(0,1)$ is added as an extra loss term.
Intuitively, the encoder learns how much noise to inject at the code level.
We use an MLP-2 in both the encoder and the decoder and optimize reconstruction loss (via binary-cross entropy) along with the KL divergence term.

\paragraph{Adversarial Autoencoders} 
Adversarial autoencoders (AAE)~\cite{Galke:2018,DBLP:journals/corr/MakhzaniSJG15} combine generative adversarial networks~\cite{DBLP:conf/nips/GoodfellowPMXWOCB14} with
autoencoders (Figure~\ref{fig:aae}). The autoencoder component reconstructs the sparse item
vectors, while the discriminator distinguishes between the generated codes and
samples from a selected prior distribution.  
Hence, the distribution of the latent code is shaped to match the prior distribution.
The latent representations learned by distinguishing the code
from a smooth prior lead to a model that is more robust to sparse input
vectors than undercomplete autoencoders because smoothness is 
the main criterion for good representations that disentangle the explanatory factors
of variation~\cite{bengio2013representation}.
Formally, we first compute \( \vz = \mlp_\text{enc}(\vx) \)
and \( \vr = \mlp_\text{dec}(\vz) \) and then update the parameters of the encoder
and the decoder with respect to the binary cross-entropy, \( \operatorname{BCE}(\vx, \vr) \).
Hence, in the regularization phase, we draw samples \( \tilde{\vz} \sim \mathcal{N}(0,
\boldsymbol{I}) \) from independent Gaussian distributions matching the size of
\( \vz \). The parameters of the discriminator \( \mlp_\text{disc} \) are then
updated, to minimize \( \log \mlp_\text{disc}(\tilde{\vz}) + \log (1 -
\mlp_\text{disc}(\vz)) \)~\cite{DBLP:conf/nips/GoodfellowPMXWOCB14}.
Finally, the parameters of the encoder are updated to
maximize \( \log \mlp_\text{disc}(\vz) \), such that the encoder is trained to
fool the discriminator. Thus, the encoder is jointly optimized for
matching the prior distribution and reconstructing the
input~\cite{DBLP:journals/corr/MakhzaniSJG15}. 
At prediction time, we perform one reconstruction step by applying one encoding
and one decoding step.

\paragraph{Conditioning on Additional Metadata as Side Information}\label{par:condition}
An advantage of auto\-enc\-oder-based recommender systems is that they
enable conditioning on side information.
This conditioning may be performed with different strategies, \eg by supplying the side information as additional input~\cite{DBLP:journals/eswa/BarbieriABZ17}.
In this work, we chose to impose the condition $\vs$ on the code level of the autoencoders. The rationale for this strategy is that the side information may help the decoder to reconstruct the item set. Such strategy is similar to the supervised case of the original work on the adversarial autoencoder that used classes as condition to reconstruct images~\cite{DBLP:journals/corr/MakhzaniSJG15}, 
and has the advantage that we can apply it in the same way to all autoencoder variants without further adaptions specific to an individual variant (such as disabling DAE's noise or modifying the VAE's KL divergence objective).
This way, the encoder still operates solely on the (partial) item set while the decoder is conditioned on side information to reconstruct the original item set:
\( \vr = \operatorname{dec}\left(
\operatorname{enc}(\vx) || \vs\right) \), where $\cdot || \cdot$ denotes concatenation. 
Furthermore, the training objective remains optimizing for the reconstruction of the item set rather than reconstruction of the side information.
In practice, we embed the documents' titles into a lower-dimensional space by using pre-trained word embeddings such as Word2Vec~\cite{DBLP:conf/nips/MikolovSCCD13}.
More precisely,
we employ a TF-IDF weighted bag of embedded words representation which has
proven
to be useful for information retrieval~\cite{DBLP:conf/gi/GalkeSS17}.
We use the same strategy for journal names as they often contain indicative words. For authors, we learn a categorical embedding from scratch: we optimize a randomly initialized embedding vector for each author during training.
In our scenarios, the side information is composed of the documents' title, journal name, and authors. We consider three cases for our experiments: (1) no conditioning, (2) conditioning on the title, and (3) conditioning on all metadata.
When multiple conditions are used, we combine them, again, via concatenation: $\vs = \vs_{\text{title}} || \vs_{\text{author}} || \vs_\text{journal}$. In our experiments, we evaluate those three cases with all the autoencoder variants described above.

\section{Experimental Apparatus}\label{sec:experiments}
In the following, we describe the datasets, our experimental procedure, baselines, and hyperparameter optimization.

\subsection{Datasets}\label{sub:ds}
We consider six datasets for our experiments.
Three datasets collect scientific publications in the domains of medicine and computer science for the citation recommendation task.
Further three datasets are in the domains of economics, politics, and news, and are used for the subject label recommendation task.
Table~\ref{tab:metadata} illustrates the metadata available in each dataset.
For each dataset, we estimate the power-law coefficient and mutual information. The former allows us to assess how skewed the distribution of documents' citations or labels' assignments is. The latter shows how informative already assigned items are for other items.
We estimate the power law coefficient $\alpha$ via maximum likelihood~\cite{newman2005power}: $\alpha = 1 + n {\left( \sum_{u \in V} \ln \frac{\textrm{deg}_u}{\textrm{deg}_{\min{}}} \right)}^{-1}$,
where $\textrm{deg}_{\min{}}$ is equal to~1. 
We compute the mutual information, \ie{} the KL Divergence of the joint distribution with respect the product of marginals: $\operatorname{MI}(X,Y) = D_{KL}(p(x,y) || p(x)p(y) )$~\cite{10.5555/1146355}. 
The distribution $p(x,y)$ models the number of documents in which two items occur together, while the marginals $p(x)$ and $p(y)$ are estimated based on the number of documents that have the item $x$. As $p(x)$ equals $p(y)$ in our case, the normalized mutual information is $\frac{MI(X,X)}{H(X)}$, with $H(X)$ being the entropy of $X$ for normalization.

\begin{table}
\centering
\small
\caption{Availability and occurrence of metadata in the datasets considered for the two recommendation tasks. Subject labels and item set occurrences are the same for the subject label datasets as the subject labels are the items to recommend (but can also be used as additional metadata for the citation tasks).}
\label{tab:metadata}
\begin{tabular}{lrrrrrr}
\toprule
Metadata field & PubMed & DBLP  & ACM   & EconBiz & IREON & Reuters \\
\midrule
Title          & 100\,\%  & 100\,\% & 100\,\% & 100\,\%   & 100\,\% & 100\,\%   \\
Author         & 100\,\%  & 100\,\% & 93\,\%  & 98\,\%    & 83\,\%  & -       \\
Abstract       & -      & 89\,\%  & 4\,\%   & -       & -     & -    \\ 
Venue/Journal title          & 100\,\%      & 83\,\%  & 100\,\% & -       & -     & -       \\
Subject labels     & 77\,\%   & -     & -     & 72\,\%       & 100\,\%     & 100\,\%       \\
Item set & 100\,\%      & 89\,\%  & 81\,\%   & 72\,\%       & 100\,\%     & 100\,\%    \\ 
\bottomrule
\end{tabular}
\end{table}

\subsubsection{Datasets for Citation Recommendation}

\begin{figure}
  \begin{subfigure}[b]{0.32\textwidth}
       \includegraphics[width=\textwidth]{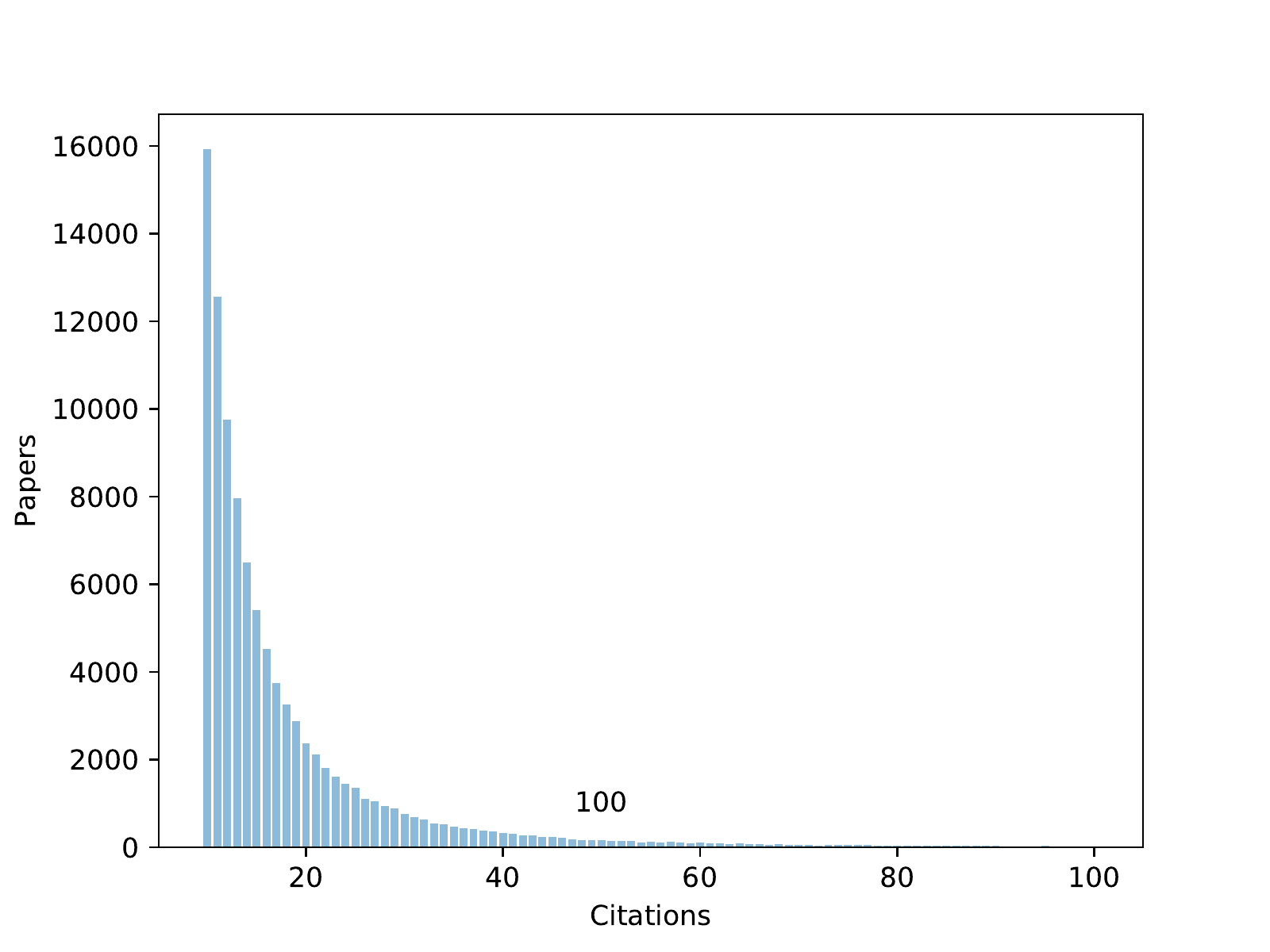}
    \end{subfigure}
    \begin{subfigure}[b]{0.32\textwidth}
        \includegraphics[width=\textwidth]{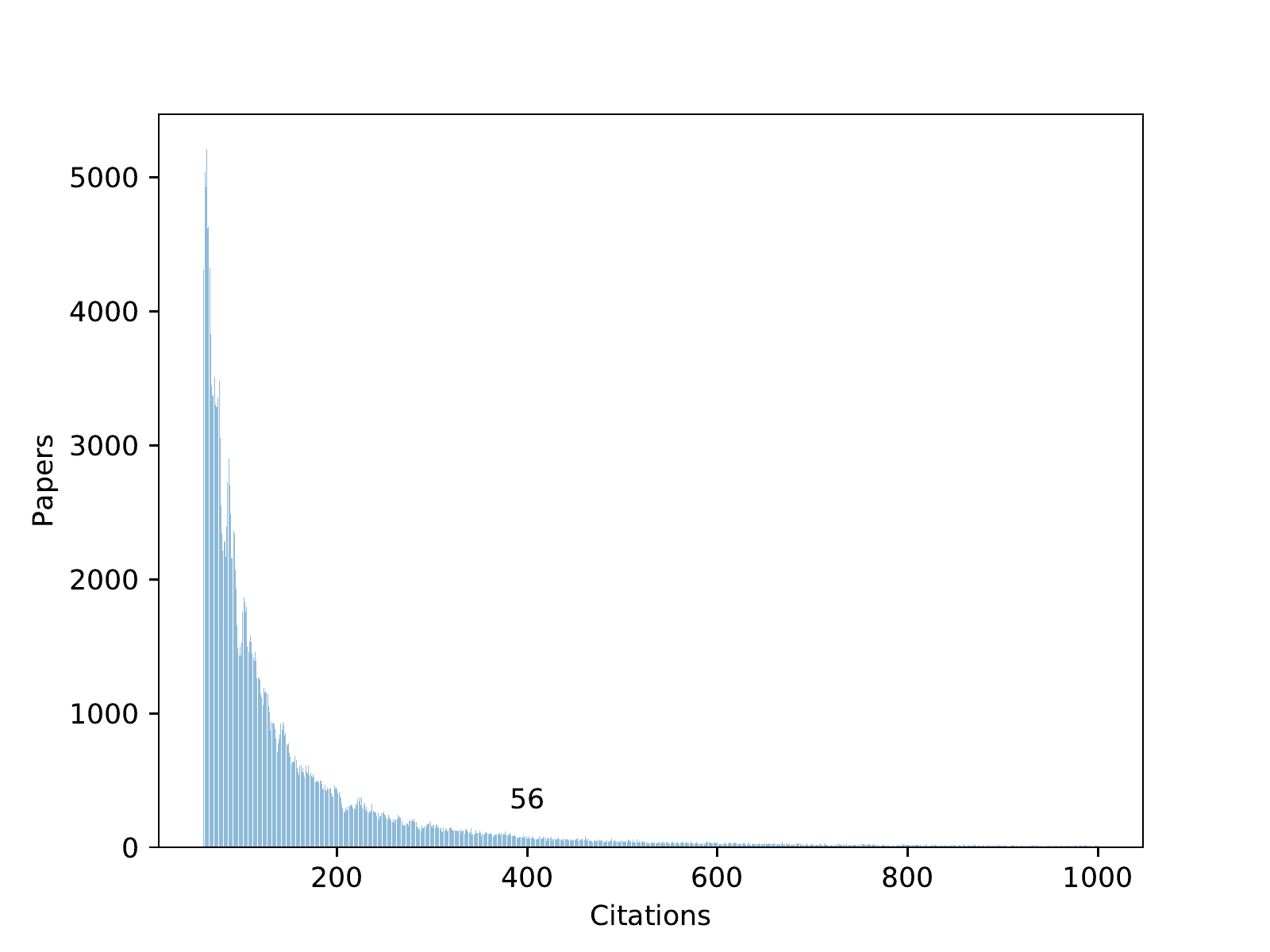}
    \end{subfigure}
    \begin{subfigure}[b]{0.32\textwidth}
        \includegraphics[width=\textwidth]{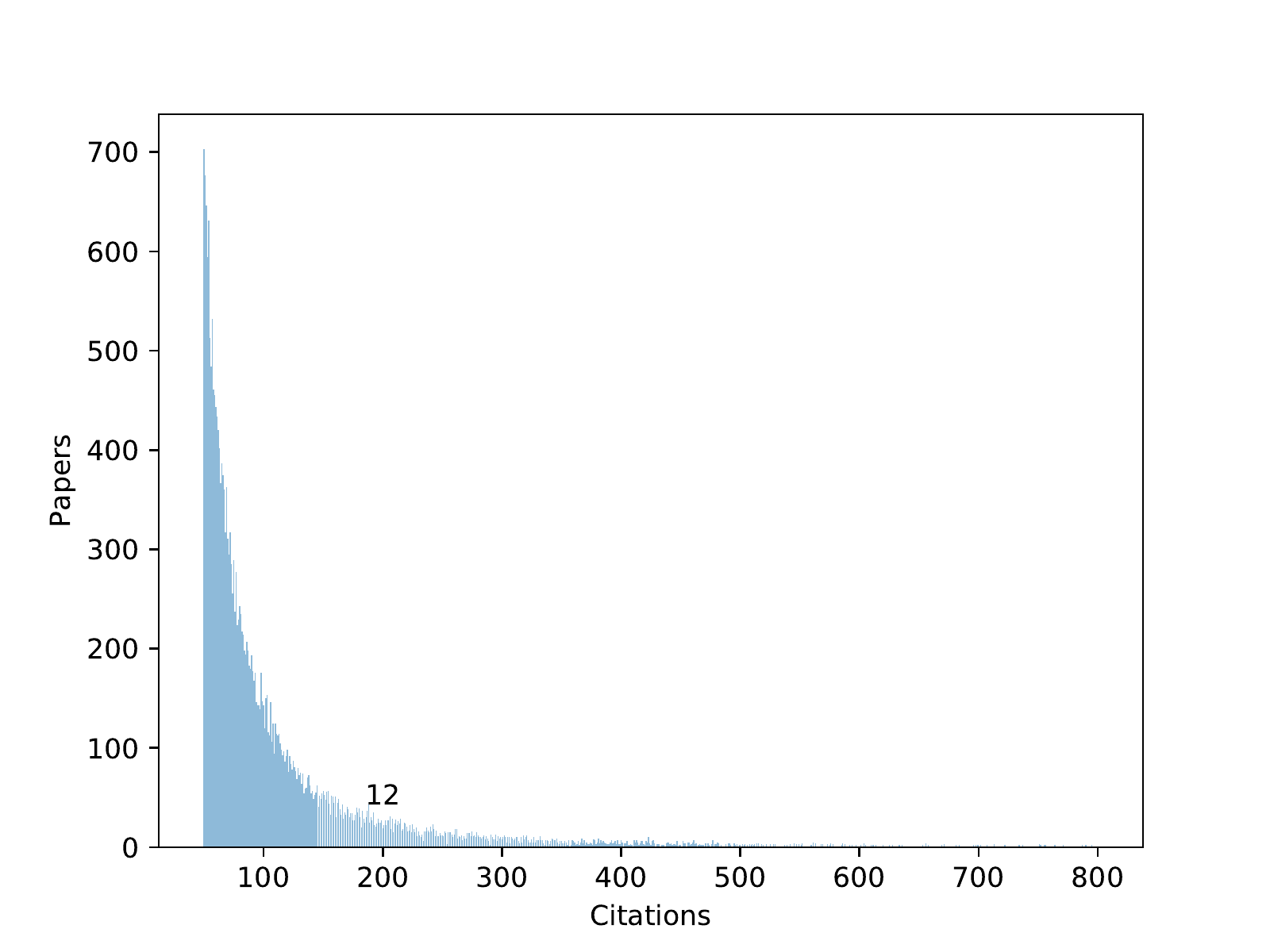}
    \end{subfigure}
     \begin{subfigure}[b]{0.32\textwidth}
       \includegraphics[width=\textwidth]{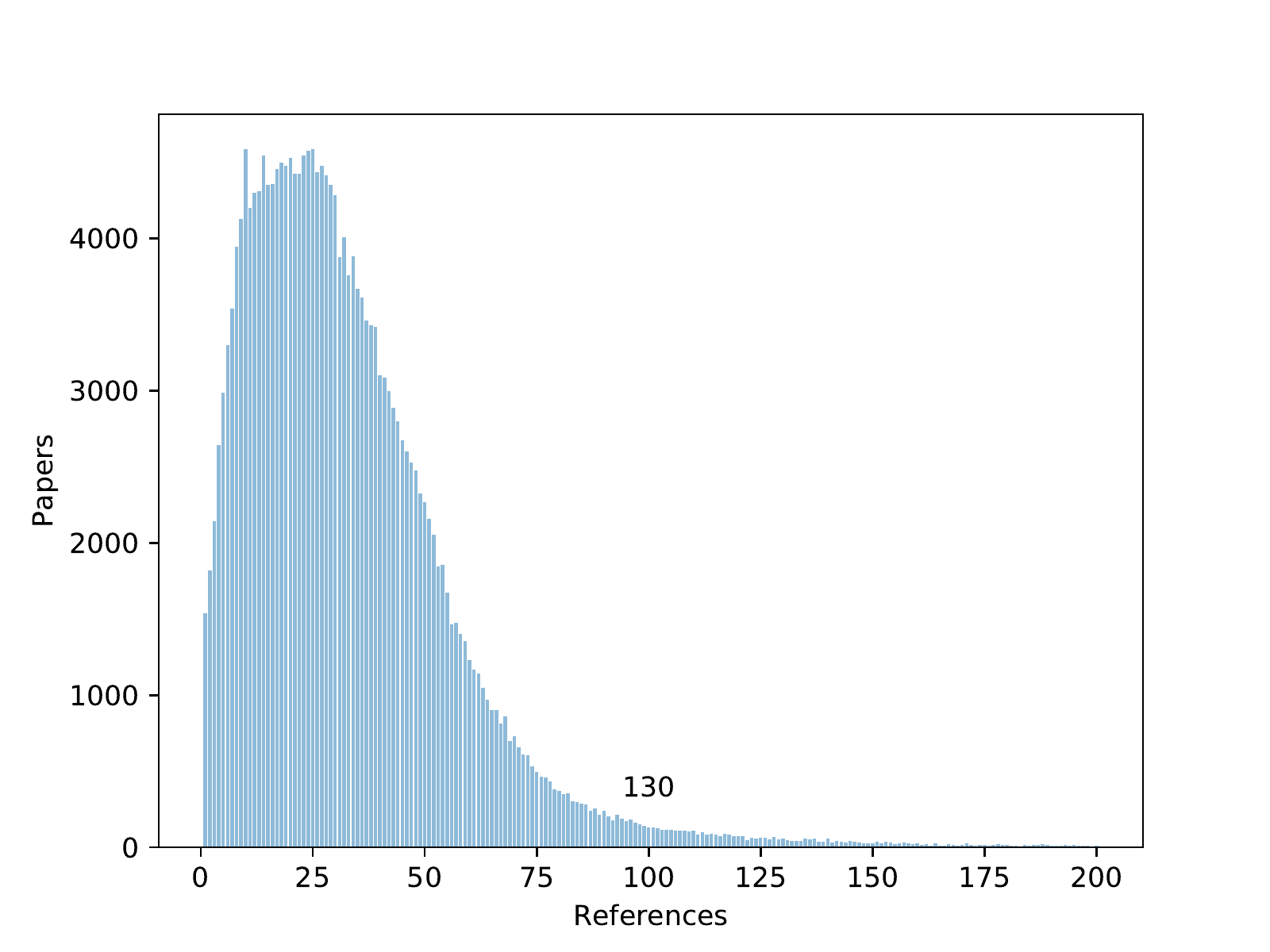}
    \end{subfigure}
      ~
    \begin{subfigure}[b]{0.32\textwidth}
        \includegraphics[width=\textwidth]{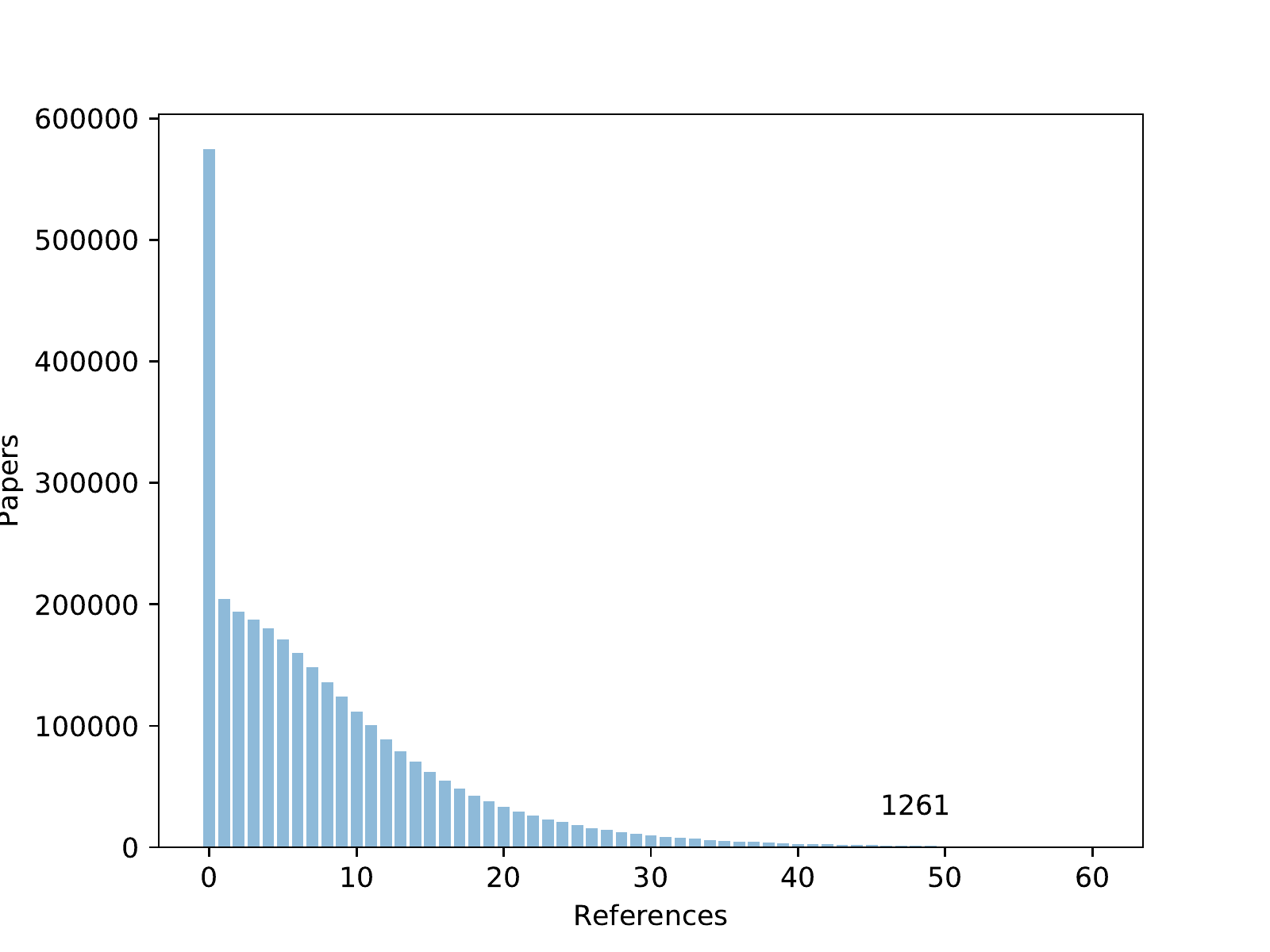}
    \end{subfigure}
    ~
    \begin{subfigure}[b]{0.32\textwidth}
        \includegraphics[width=\textwidth]{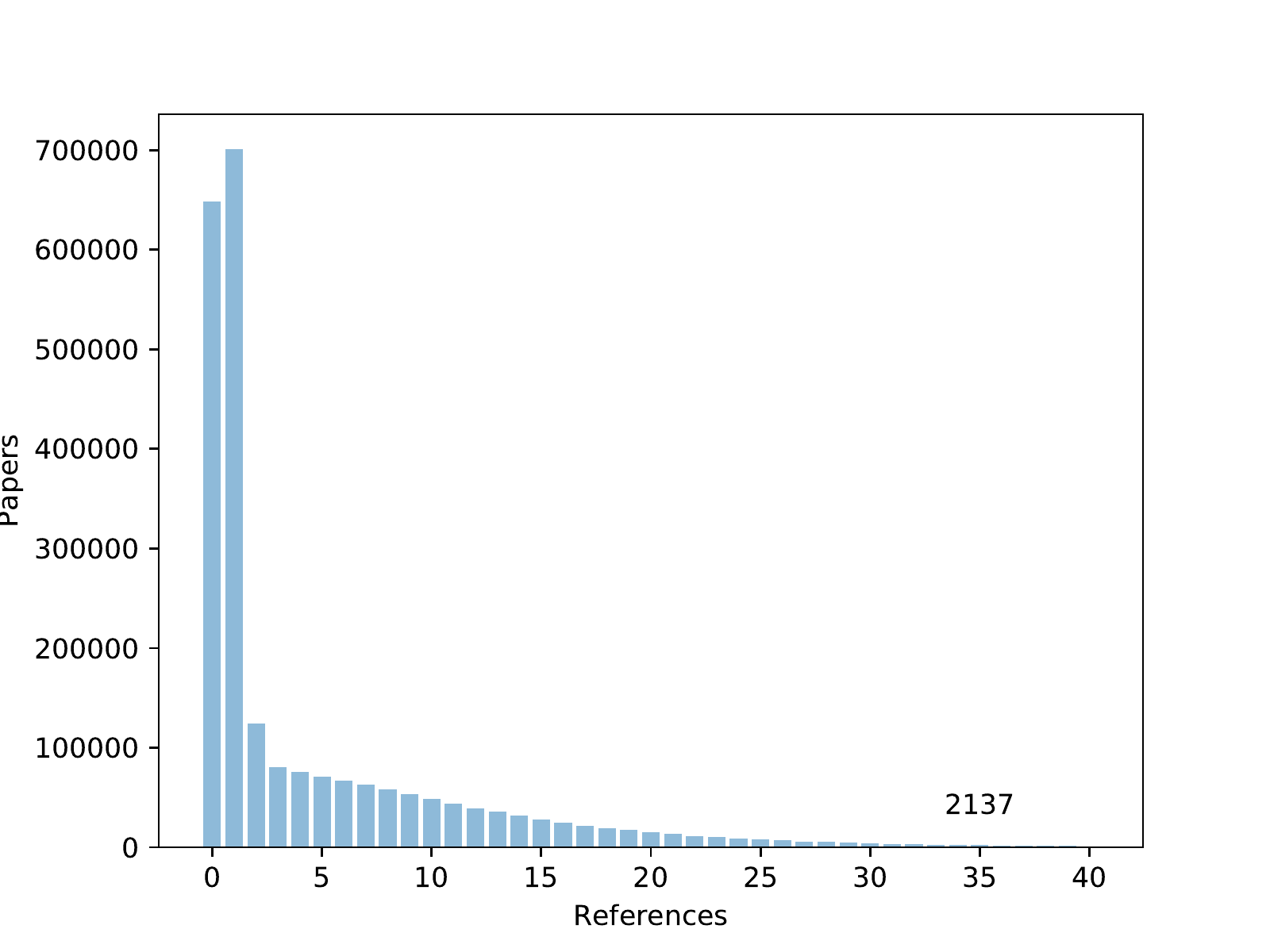}
    \end{subfigure}
  \caption{Documents by citations for PubMed (top-left), DBLP (top-center) and ACM (top-right) datasets,
 and by the number of cited documents for PubMed (bottom-left), DBLP (bottom-center) and ACM (bottom-right).
}
  \label{fig:docs_stats}
\end{figure}

\paragraph{PubMed Citation Dataset}
The CITREC\footnote{\url{https://www.isg.uni-konstanz.de/projects/citrec/}} PubMed
citation dataset~\cite{gipp2015citrec} consists of 7,546,982 citations.
The dataset comprises $224,092$ distinct citing documents published between 1928
and 2011 and $2,896,764$ distinct cited documents.
Each document has an identifier, article title, title of the journal where is published, list of authors, Medical Subject Headings (MeSH)\footnote{\url{https://www.nlm.nih.gov/mesh/}} labels, and the publication year.
The documents are cited between $1$ and $3,247$ times with a median of $1$
and a mean of $2.61$ (SD\@: $6.71$).
The citing documents hold on average $33.68$ (standard deviation, SD\@: $27.49$) citations to other documents (minimum: $1$, maximum: $2,242$) with a median of $29$.
This dataset, like all the others but Reuters, seems to follow a power-law distribution, which is typical for citation networks (Figure \ref{fig:docs_stats}).
The $\alpha$ coefficient for the PubMed dataset is $1.47$ for citations and $1.30$ for the number of cited documents.
The normalized mutual information for PubMed is $0.5996$.

\paragraph{DLBP Citation Dataset}
The DLBP Citation Network\footnote{\label{note:aminer}\url{https://aminer.org/citation}}~\cite{Tang2008} includes $25,166,994$ citations.
The dataset comprises $3,079,007$ distinct citing documents published between 1936 and 2018 and $1,985,921$ distinct cited documents.
Each document has an identifier, title, publication venue, authors, number of citations, publication date, and optionally the abstract. 
The documents are cited between $1$ and $16,229$ times with a median of $4$ and a mean of $12.67$ (SD\@: $56.17$).
The citing documents hold on average $8.17$ (SD\@: $9.71$) citations to other documents (minimum: $0$, maximum: $1,532$) with a median of $6$.
The $\alpha$ coefficient is $1.58$ for citations and $1.28$ for the number of cited documents. 
The normalized mutual information is $0.5407$.

\paragraph{ACM Citation Dataset}
The ACM Citation Network\textsuperscript{\ref{note:aminer}}
~\cite{Tang2008} contains $11,344,141$ citations. 
The dataset comprises $2,385,066$ distinct citing documents published between 1936 and 2016 and $2,631,128$ distinct cited documents. 
Each document is characterized by its identifier, title, publication venue, authors, publication date, and optionally the abstract. 
The documents are cited between $0$ and $810$ times with a median of $1$ and a mean of $4.76$ (SD\@: $7.74$).
The citing documents hold on average $4.31$ (SD\@: $580.96$) citations to other documents (minimum: $1$, maximum: $938,039$) with a median of $1$.
The $\alpha$ coefficient is $1.53$ for citations and $1.32$ for the number of cited documents, while the normalized mutual information is $0.5282$.

\subsubsection{Datasets for Subject Labels Recommendation}

\begin{figure}
  \begin{subfigure}[b]{0.32\textwidth}
       \includegraphics[width=\textwidth]{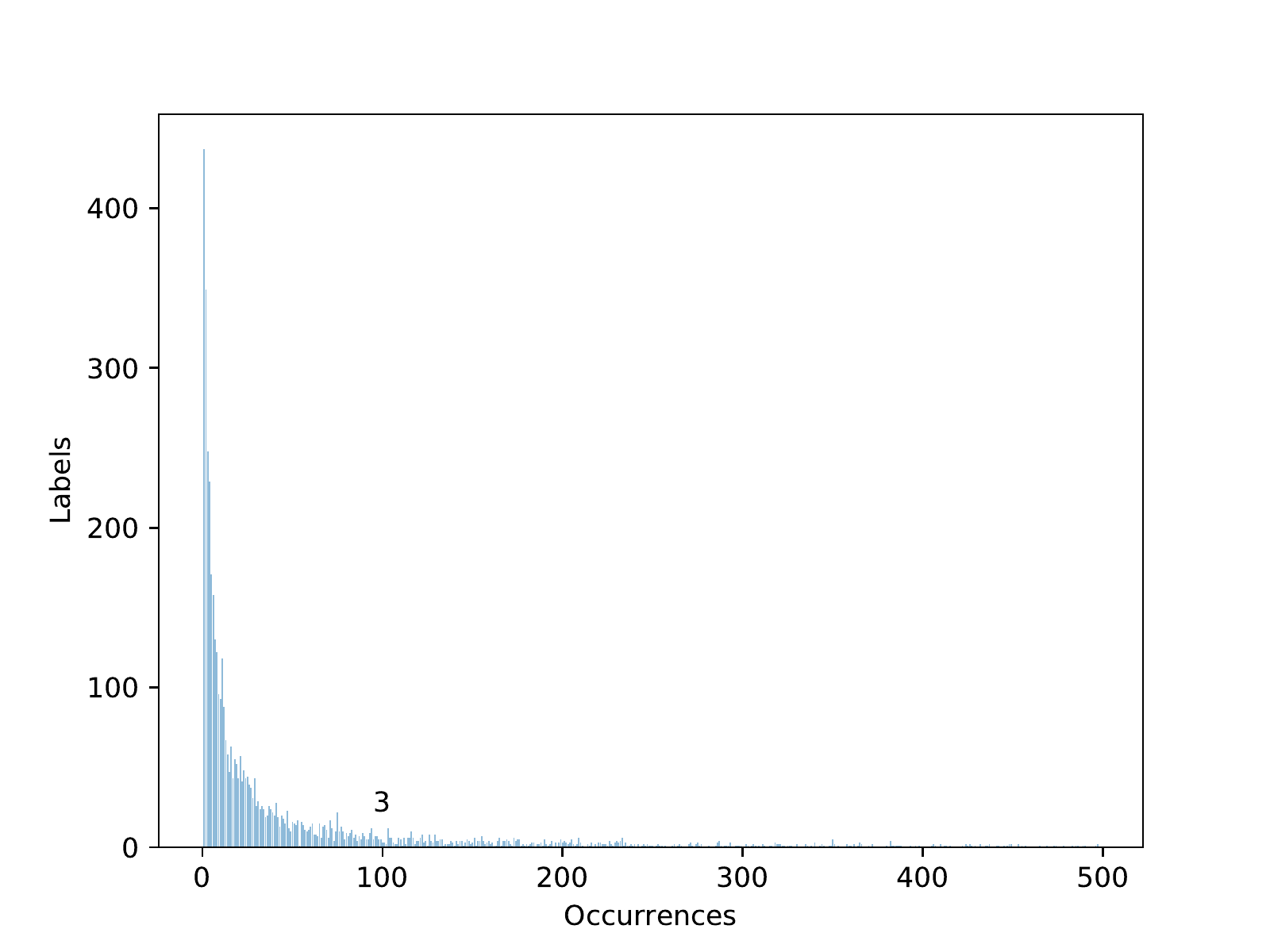}
    \end{subfigure}
    \begin{subfigure}[b]{0.32\textwidth}
        \includegraphics[width=\textwidth]{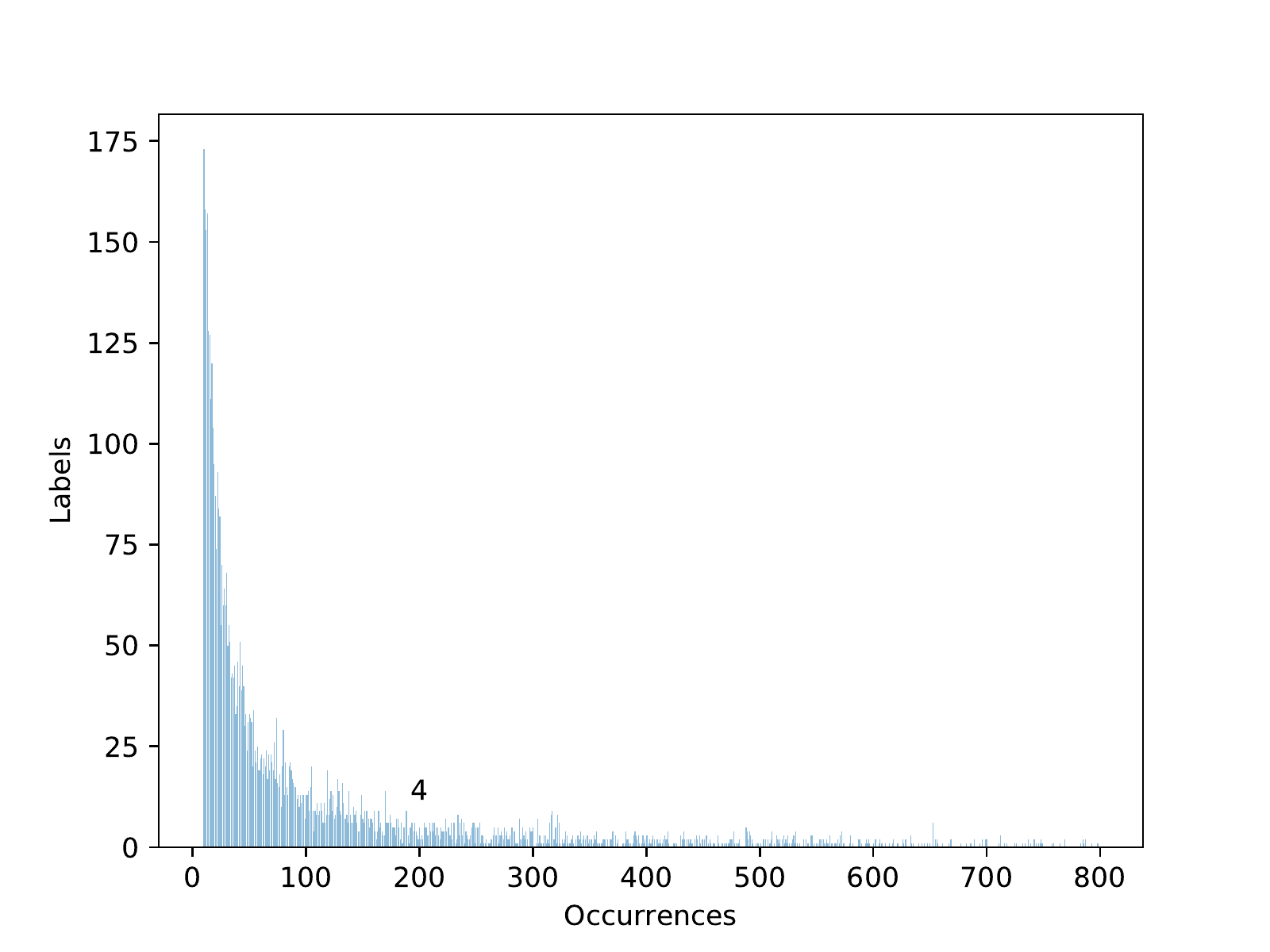}
    \end{subfigure}
    \begin{subfigure}[b]{0.32\textwidth}
        \includegraphics[width=\textwidth]{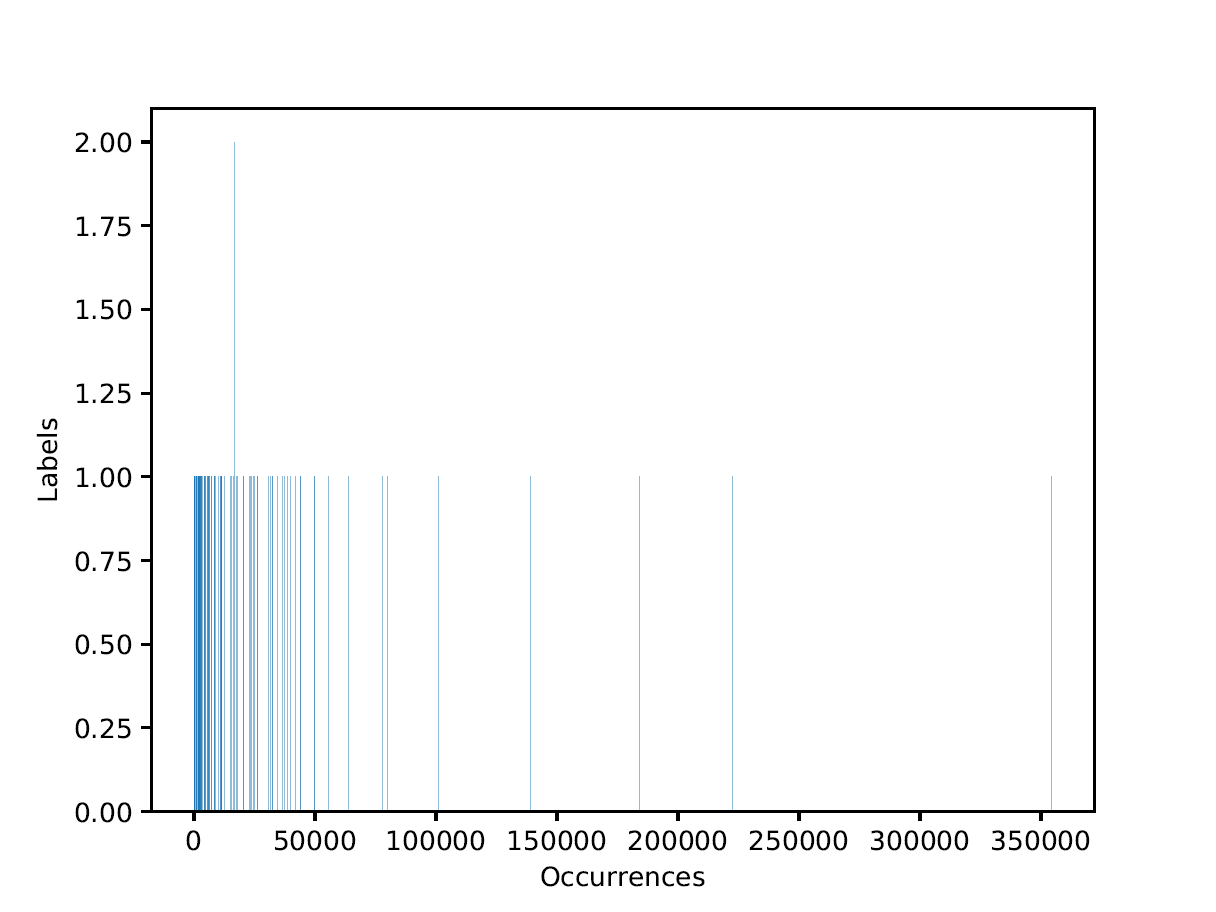}
    \end{subfigure}
    
    \begin{subfigure}[b]{0.32\textwidth}
       \includegraphics[width=\textwidth]{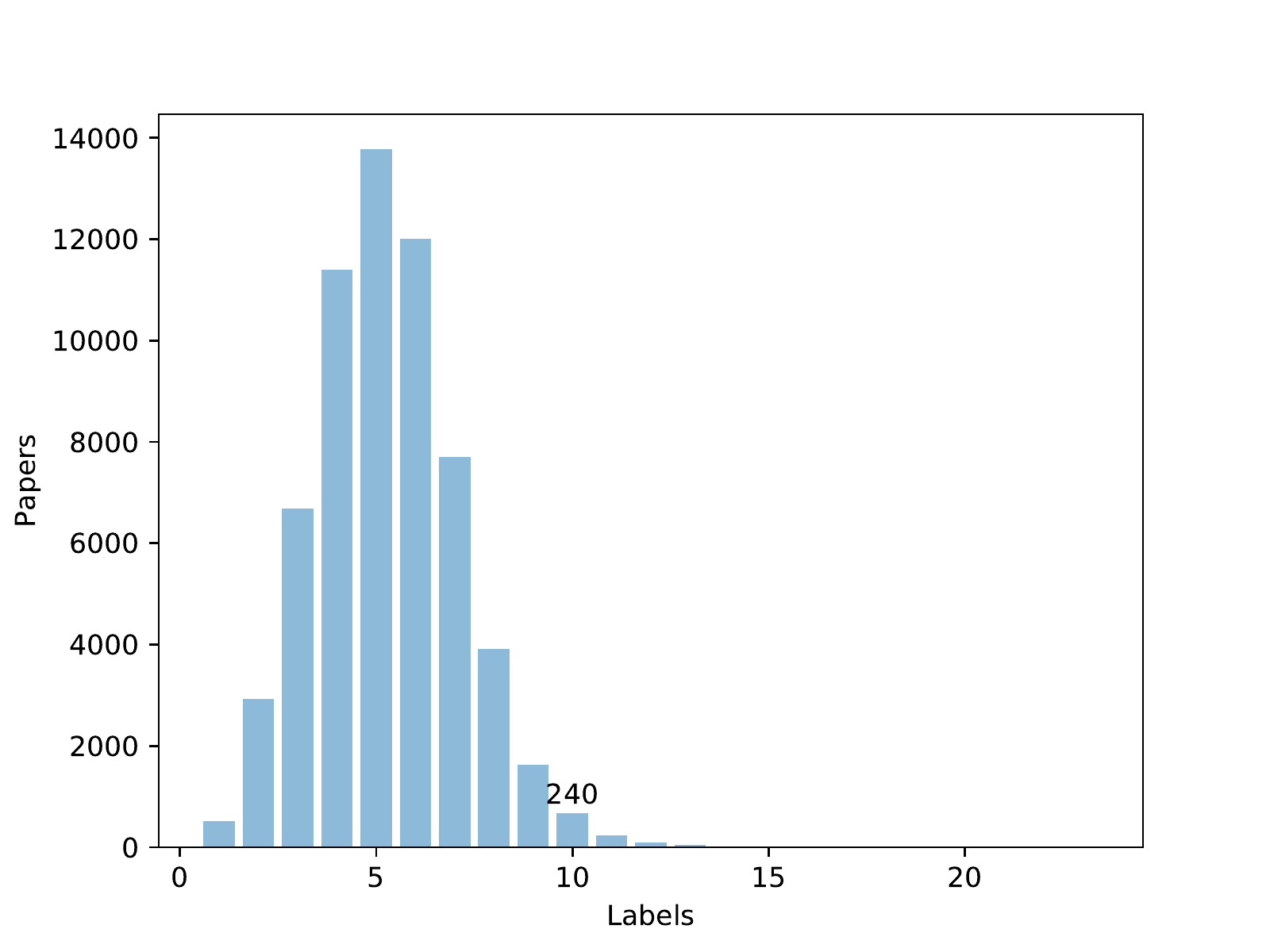}
    \end{subfigure}
    \begin{subfigure}[b]{0.32\textwidth}
        \includegraphics[width=\textwidth]{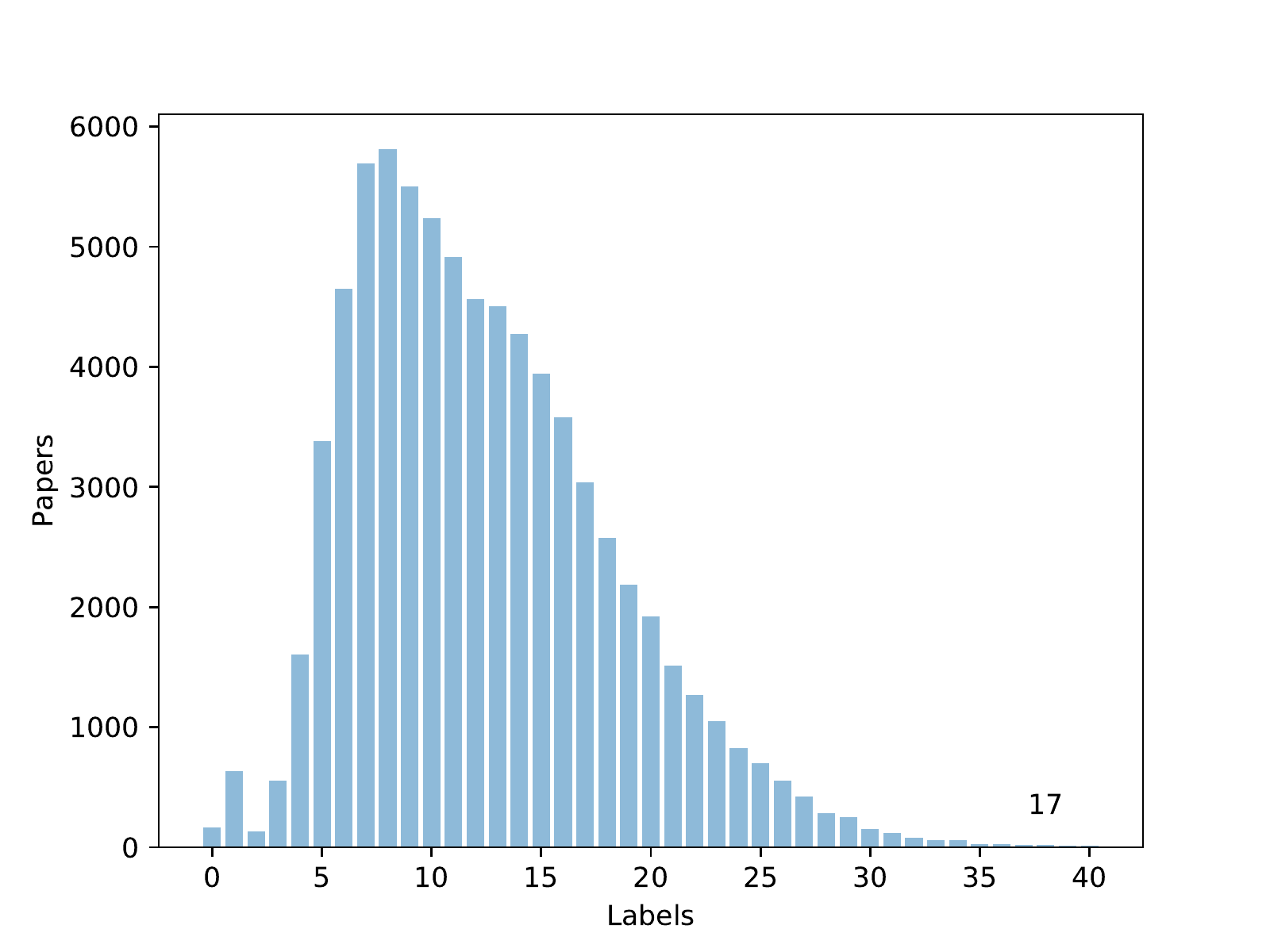}
    \end{subfigure}
    \begin{subfigure}[b]{0.32\textwidth}
        \includegraphics[width=\textwidth]{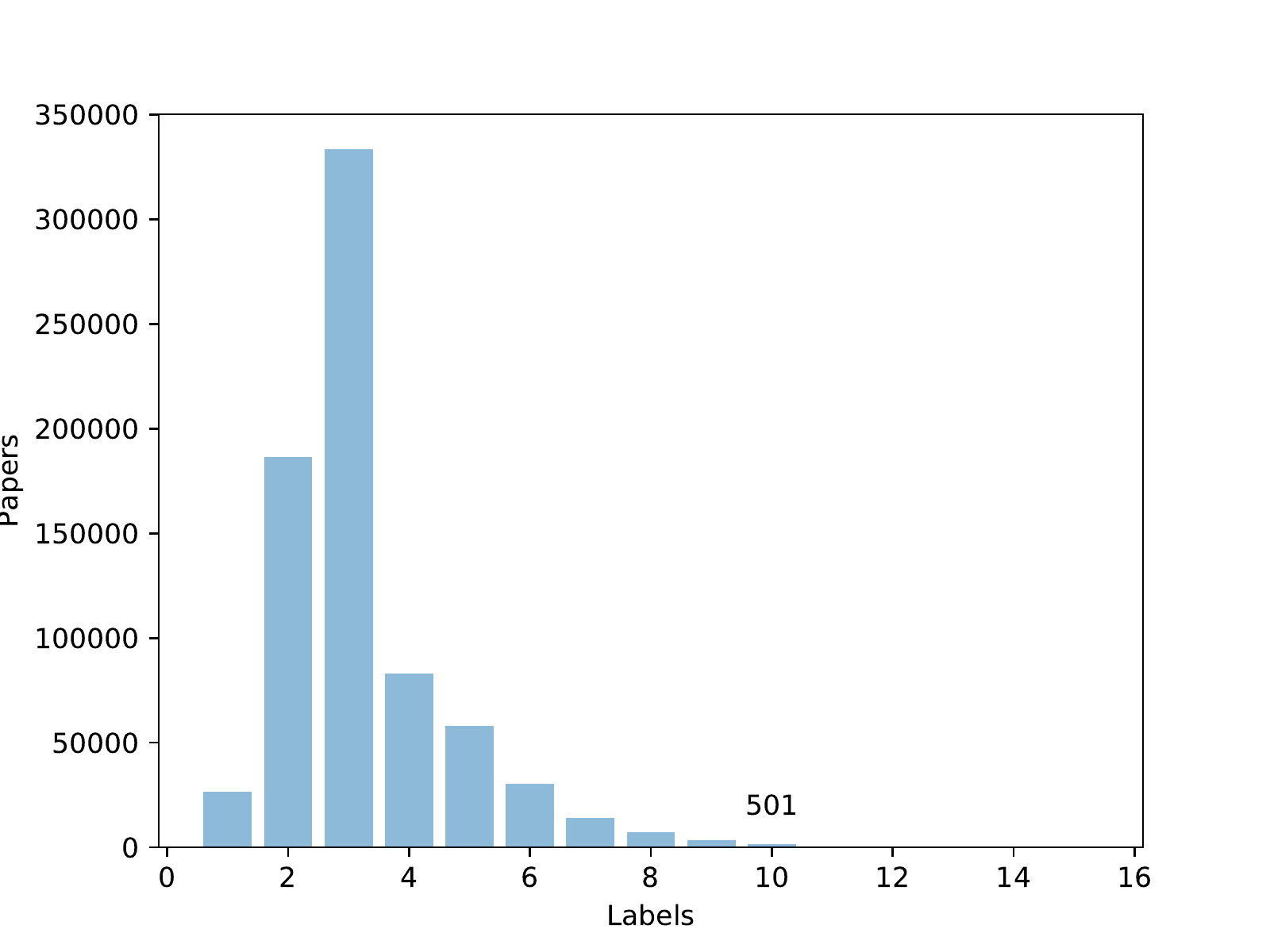}
    \end{subfigure}
  \caption{Subject labels by the number of times they have been assigned to documents (occurrence) for the Economics (top-left), IREON (top-center) and Reuters (top-right) datasets,
and documents by the number of labels they have been assigned for EconBiz (bottom-left), IREON (bottom-center) and Reuters (bottom-right).
}
  \label{fig:label_stats}
\end{figure}

\paragraph{EconBiz Subject Labels}
The EconBiz dataset\footnote{\url{https://www.kaggle.com/hsrobo/titlebased-semantic-subject-indexing}}, provided by ZBW --- Leibniz Information Centre for Economics,
consists of $61,619$ documents with label annotations from professional subject
indexers~\cite{DBLP:conf/kcap/GalkeMSBS17,DBLP:conf/kcap/Grosse-BoltingN15}. 
The $4,669$ assigned labels are a subset of the controlled vocabulary Standardthesaurus
Wirtschaft\footnote{\url{http://zbw.eu/stw/version/latest/about}}, a professional thesaurus for economics. 
Every document has an identifier, title, authors, language label(s), subject labels, and publication year, as well as optionally the publisher, publication country, and series.
The number of documents to which a subject label is assigned ranges between $1$ and $13,925$ with a mean of $69$ (SD\@: $316$) and a median of $14$. 
The label annotations of a document range between $1$ and $23$ with a mean of $5.24$ (SD\@: $1.83$) and a median of $5$ labels.
As for citations, the subject label datasets follow a power law distribution, except Reuters (see Figure \ref{fig:label_stats}).
The $\alpha$ coefficient for the EconBiz dataset is $1.96$ for the label occurrences and $1.19$ for the number of assigned labels. 
The normalized mutual information is $0.2970$.

\paragraph{IREON Subject Labels}
The political sciences dataset IREON has \( 76,359 \) documents provided by the German Information Network for International Relations and Area Studies\footnote{\url{http://www.fiv-iblk.de/eindex.htm}}. Each document holds an identifier, title, authors, language label, subject labels, and publication year.
The \( 10,440 \) subject labels assigned to the papers are taken from the thesaurus for International Relations and Area
Studies\footnote{\url{http://www.fiv-iblk.de/information/information_thesaurus.htm}}.
The number of documents to which a label is assigned ranges between $1$ and $13,895$ with a mean of $90.68$ (SD\@:~$338.63$) and a median of $13$. 
The label annotations of a document range between $0$ and $70$ with a mean of $12.40$ (SD\@: $5.91$) and a median of $11$.
The $\alpha$ coefficient is $1.94$ for the label occurrences and $1.21$ for the number of assigned labels, while the normalized mutual information is $0.1977$.

\paragraph{Reuters Subject Labels}
The Reuters RCV1-v2 dataset contains \( 744,693 \) news articles and a thesaurus providing a set of $104$ labels~\cite{lewis04}.
Every document includes an identifier, title, subject labels, and publication date.
The number of documents to which a label is assigned ranges between $5$ and $354,437$ with mean $23,217.35$ (SD\@: $47,313.01$) and median $7,315$. 
The label annotations of a document range between $1$ and $17$ with a mean of $3.24$ (SD\@: $1.42$) and a  median of $3$ labels.
Its label occurrence does not follow a power-law distribution (Figure~\ref{fig:label_stats}).
The $\alpha$ coefficient is $148.15$ for label occurrence and $1.14$ for the number of assigned labels.
The normalized mutual information is $0.3207$.

\subsection{Experimental Procedure}\label{sub:procedure}

\begin{figure}[!h]
  \begin{subfigure}{0.32\textwidth}
       \includegraphics[width=\textwidth]{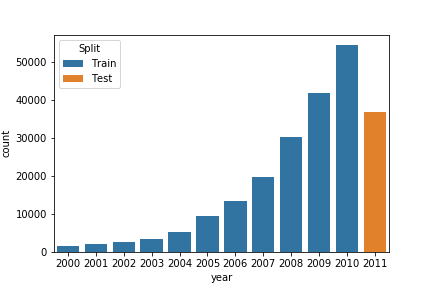}
    \end{subfigure}
    \begin{subfigure}{0.32\textwidth}
        \includegraphics[width=\textwidth]{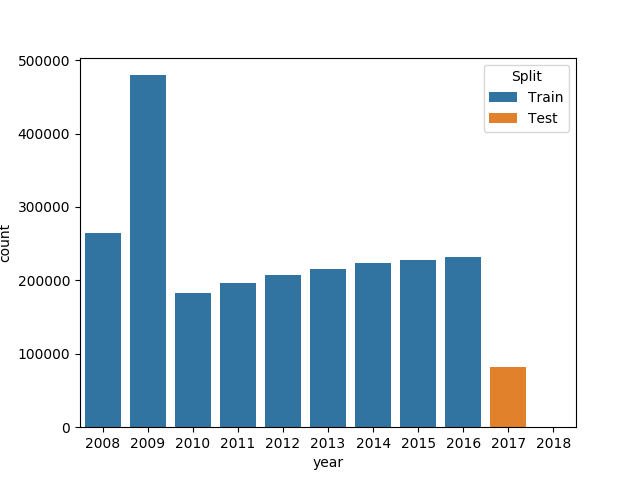}
    \end{subfigure}
    \begin{subfigure}{0.32\textwidth}
        \includegraphics[width=\textwidth]{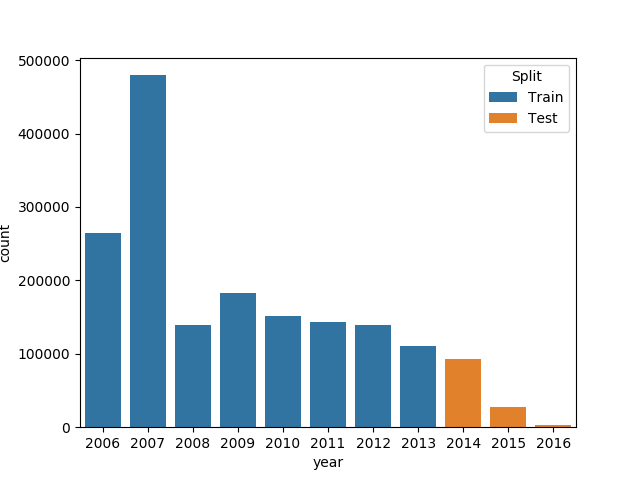}
    \end{subfigure}
    
    \begin{subfigure}{0.32\textwidth}
       \includegraphics[width=\textwidth]{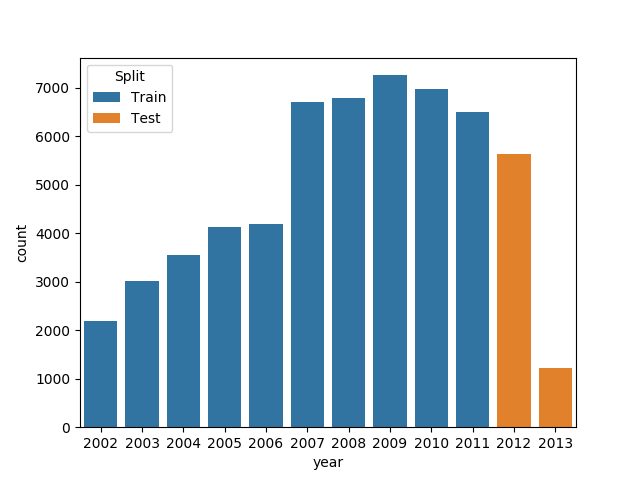}
    \end{subfigure}
    \begin{subfigure}{0.32\textwidth}
        \includegraphics[width=\textwidth]{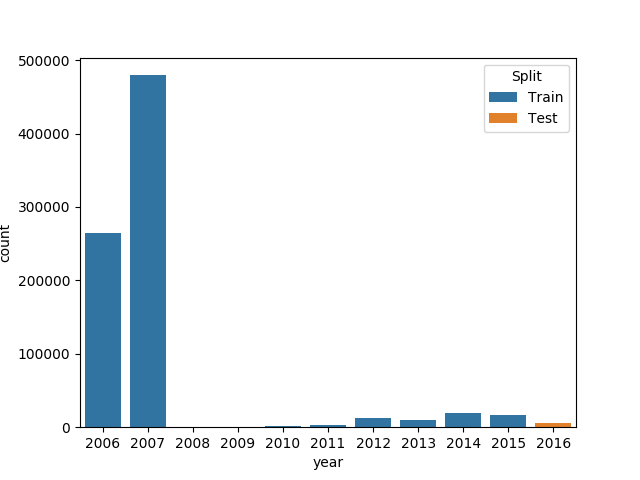}
    \end{subfigure}
    \begin{subfigure}{0.32\textwidth}
        \includegraphics[width=\textwidth]{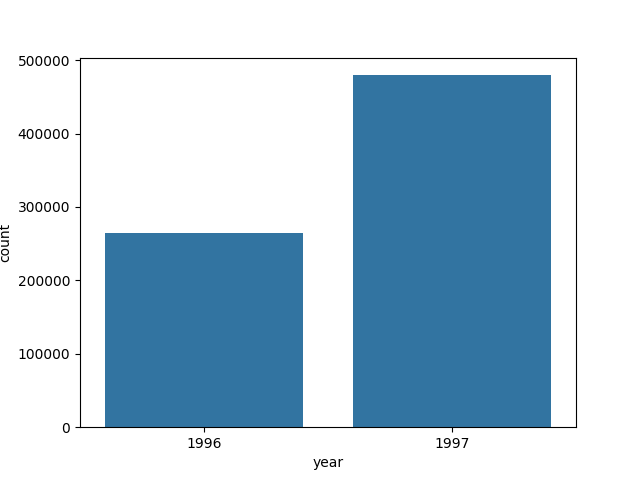}
    \end{subfigure}
  \caption{Count of documents by publication year with
  the split in training and test set for PubMed (top left), DBLP (top center) and ACM (top right) EconBiz (bottom left), IREON (bottom center), and Reuters (bottom right) datasets, starting from 2000, 2008, 2006, 2002, and 2006, respectively, for the first five datasets. For Reuters training and test set are randomly generated.}\label{fig:pmcc_by_year}
\end{figure}

\paragraph{Train-Test Split of the Datasets along the Time Axis} To simulate a real-world citation prediction
setting, we split the data on the time axis of the citing documents. This
resembles the natural constraint that publications cannot cite other
publications that do not exist yet. Given a specific publication year \(T\), we
ensure that the training set, \( \mathbb{D}_\train \), consists of all documents that were
published earlier than year \(T\), and use the remaining documents as test data,
\(\mathbb{D}_\test \).
Figure~\ref{fig:pmcc_by_year} shows the distribution of documents over the years
along with the split into training set and test set for PubMed, DBLP, and ACM. 
Regarding our evaluation, we select the year 2011 for PubMed, 2017 and 2018 for DBLP, and 2014 to 2016 for ACM to obtain a 90:10 ratio between training and test documents.

We also conduct the split between the training and test set along the time axis for the EconBiz and IREON datasets in the subject labeling scenario (Figure~\ref{fig:pmcc_by_year}). 
This is challenging because label annotations suffer
from concept drift over time~\cite{DBLP:conf/jcdl/ToepferS17}. We use the years 2012 and 2013 as test
documents for EconBiz and the year 2016 for IREON to obtain a 90:10 train-test ratio, similar to the citation task. 
Since Reuters contained just two years (1996 and 1997), a time split would have generated too few documents from 1996 to obtain such a ratio. 
Here, we randomly select 10\,\% of documents for the test set, regardless of the year.

\paragraph{Preprocessing and Dataset Pruning as a Controlled Variable}
For preprocessing the datasets, we conduct the following three steps:
%
First, we build a vocabulary on the training set with items cited/assigned more than \( k \) times.
 Second, we filter both the training and test set and retain only items from the vocabulary.
 Finally, we remove documents with fewer than two of the vocabulary items in their item set.

\begin{table}[!h]
  \small
  \centering
  \caption{Characteristics of the citation datasets with respect to different selected pruning thresholds on the minimum item occurrence.}\label{tab:citation:pruning}
  \begin{tabular}{lrrrrrrr}
\toprule
Dataset & Pruning &  \begin{tabular}[x]{@{}r@{}}Cited \\docs \end{tabular}&  Citations &  Documents & Density & $\alpha$ & MI \\
\midrule
\multirow{4}{*}{PubMed} 
& 20 &    20,270 &     878,359 &   121,374 &  0.000357 & 1.5870 & 0.4265\\
& 30 &     8,906 &     568,563 &    96,980 &  0.000658 & 1.6465 & 0.3958\\
& 40 &     4,939 &     413,746 &    79,830 &  0.001049 & 1.7090 & 0.3737\\
& 50 &     3,185 &     324,693 &    67,703 &  0.001506 & 1.7755 & 0.3587\\
\midrule
\multirow{4}{*}{DBLP} & 20 &   251,405 &   16,340,121 &  1,955,132 &  0.000033 & 1.5960 & 0.4879\\
& 30 &   157,203 &   13,977,243 &  1,839,181 &  0.000048 & 1.6046 & 0.4750\\
& 40 &   109,469 &   12,271,346 &  1,742,180 &  0.000064 & 1.6127 & 0.4658\\
& 50 &    81,817 &   10,991,096 &  1,660,462 &  0.000081 & 1.6206 & 0.4591\\
\midrule
\multirow{4}{*}{ACM}
& 20 &    78,805 &    5,590,751 &   786,216 &  0.000090 & 1.5840 & 0.4650\\
& 30 &    46,782 &    4,752,086 &   751,981 &  0.000135 & 1.6099 & 0.4521\\
& 40 &    31,780 &    4,189,331 &   725,635 &  0.000182 & 1.6354 & 0.4435\\
& 50 &    23,177 &    3,767,585 &   702,158 &  0.000232 & 1.6610 & 0.4374\\
\bottomrule
\end{tabular}

\end{table}

\begin{table}[!h]
  \small
  \centering
  \caption{Characteristics of the subject label datasets with respect to different selected pruning thresholds on minimum item occurrence.}\label{tab:label:pruning}
  \begin{tabular}{lrrrrrrr}
\toprule
Dataset & Pruning &  Classes &  Labels &  Documents & Density & $\alpha$ & MI\\
\midrule
\multirow{4}{*}{EconBiz}
& 1  &     4,568 &     323,670 &    61,104 &  0.001160 & 1.9612 & 0.2970\\
& 5  &     3,259 &     320,048 &    60,983 &  0.001610 & 2.0018 & 0.2917\\
& 10 &     2,597 &     314,738 &    60,778 &  0.001994 & 2.0483 & 0.2857\\
& 20 &     1,924 &     303,693 &    60,272 &  0.002619 & 2.1379 & 0.2768\\
\midrule
\multirow{4}{*}{IREON} 
& 1  &    10,324 &     945,888 &    75,558 &  0.001213 & 1.9382 & 0.1977\\
& 5  &     6,971 &     938,677 &    75,555 &  0.001782 & 1.9644 & 0.1928\\
& 10 &     5,612 &     928,701 &    75,551 &  0.002190 & 1.9930 & 0.1881\\
& 20 &     4,304 &     909,156 &    75,535 &  0.002797 & 2.0463 & 0.1809\\
\midrule
\multirow{4}{*}{Reuters}
& 1  &      104 &    2,340,132 &   744,693 &  0.030215 & 148.15 & 0.3207\\
& 5  &      104 &    2,340,132 &   744,693 &  0.030215 & 148.15 & 0.3207\\
& 10 &      103 &    2,340,127 &   744,693 &  0.030509 & 146.71 & 0.3207\\
& 20 &      103 &    2,340,127 &   744,693 &  0.030509 & 146.71 & 0.3207\\
\bottomrule
\end{tabular}

\end{table}

The pruning threshold \( k \) is crucial since it affects both the number of
considered items as well as the number of documents.
Thus, we control \( k \) in our experiments and evaluate the models' performance with respect to its different values.
Table~\ref{tab:citation:pruning} shows the characteristics of PubMed, DBLP, and ACM with respect to \( k \), while Table~\ref{tab:label:pruning} illustrates the same for EconBiz, IREON, and Reuters.

\subsection{Baselines}\label{sec:baselines}

The MLP is not only exploited as a building block of autoencoders, but also as an additional baseline model.
Furthermore, we use two competitive baselines based on item co-occurrence and singular value decomposition. 

\paragraph{Multi-layer Perceptron (MLP)}
As neural baseline, we use the MLP introduced in \Secref{sec:models}, which only operates on the documents' metadata.
In this case, we optimize binary cross-entropy \( \operatorname{BCE}( \vx,
\operatorname{MLP-2}(\vs) ) \), where the title and other metadata \( \vs \) are used as input and
citations or subject labels \( \vx \) as target outputs.
With the purpose of a fair comparison, we operate on the same TF-IDF weighted embedded bag-of-words representation~\cite{DBLP:conf/gi/GalkeSS17}, $\vs$, as we have also used to condition the decoder of the autoencoder variants (see \Secref{par:condition}).

\paragraph{Singular Value Decomposition} Singular value decomposition (SVD)
 factorizes the co-occurrence matrix of items \( \boldsymbol{X}^T \cdot \boldsymbol{X} \).
Caragea et al. successfully used SVD for citation
recommendation~\cite{caragea2013can}. 
We include an extended version of SVD in our comparison, which can incorporate title information~\cite{Galke:2018}. 
We concatenate the textual
features as TF-IDF weighted bag-of-words with the items and perform singular value
decomposition on the resulting matrix.
To obtain predictions, we only use those
indices of the reconstructed matrix that are associated with items.

\paragraph{Item Co-Occurrence}\label{methods:baseline:cocit} As a non-parametric
yet strong baseline we consider the co-citation
score~\cite{DBLP:journals/jasis/Small73}, purely based on item co-occurrence.
The rationale is that two papers, which have been cited more often
together in the past, are more likely to be cited together in the future than
papers that were less often cited together.
Given training data, \(\boldsymbol{X}_\train \), we compute the full item
co-occurrence matrix \(\boldsymbol{C} = {\boldsymbol{X}_\train}^T \cdot
\boldsymbol{X}_\train \in \mathbb{R}^{n \times n}\). At prediction time,
we obtain the scores by aggregating the co-occurrence values via matrix
multiplication \(\boldsymbol{X}_\test \cdot \boldsymbol{C}\). On the diagonal of
\( \boldsymbol{C} \), the occurrence count of each item is retained to
model the prior probability.

\subsection{Hyperparameter Optimization} 

The hyperparameters are selected by conducting preliminary experiments on Pub\-Med by considering only items that appear $50$ or more times in the whole corpus. 
We chose this scenario because this aggressive pruning results in numbers of distinct items and documents that are similar to the
ones of the subject label recommendation datasets. Considering the MLP-modules, we conducted a grid
search with hidden layer sizes between $50$ and $1,000$, an initial learning
rates between \(0.01\) and \(0.00005 \), activation functions Tanh,
ReLU~\cite{DBLP:conf/icml/NairH10}, SELU~\cite{DBLP:conf/nips/KlambauerUMH17}
along with dropout~\cite{DBLP:journals/jmlr/SrivastavaHKSS14} (or alpha-dropout in case of SELUs) probabilities between
\(0.1\) and \(0.5\), and as optimization algorithms a standard stochastic gradient descent and
Adam~\cite{DBLP:journals/corr/KingmaB14}. 
For the autoencoder-based models, we
considered code sizes between $10$ and $500$, but only if the size was
smaller than the hidden layer sizes of the MLP modules. 
In the case of adversarial
autoencoders, we experimented with Gaussian, Bernoulli, and Multinomial prior
distributions, and with linear, sigmoid, and softmax activation on the code layer,
respectively.

Although a certain set of hyperparameters may
perform better in a specific scenario, we select the following, most
robust, hyperparameters:
hidden layer sizes of $100$ with ReLU~\cite{DBLP:conf/icml/NairH10}
nonlinearities and drop probabilities of \( 0.2 \) after each hidden
layer. 
The optimization is carried out by Adam~\cite{DBLP:journals/corr/KingmaB14}
with initial learning rate \( 0.001 \). 
The autoencoder variants use a code size of $50$. 
We further select a
Gaussian prior distribution for the adversarial autoencoder.
For SVD, we consecutively increased the number of singular values up to $1,000$.
Higher amounts of singular values decreased the performance.
We keep this set of hyperparameters across all models and subsequent experiments to ensure a reliable comparison of the models.

\subsection{Performance Measures} 

For the evaluation, certain items were omitted on purpose in the test set.
For each document, the models ought to predict the omitted items as well as possible. 
Thus, we choose the mean reciprocal rank (MRR) as our evaluation metric~\cite{Craswell2009}. We are given a set of predictions, \(\boldsymbol{X}_\text{pred} \), for the test set, \( \boldsymbol{\tilde X}_\test \). 
For each row, we compute the reciprocal rank of the missing items \( \boldsymbol{x}_\test - \boldsymbol{\tilde x}_\test \).
The reciprocal rank corresponds to one divided by the position of the first omitted item in the sorted set of predictions, \(\boldsymbol{x}_\text{pred} \).
We then average over all documents of the test set to obtain the mean reciprocal rank.
To alleviate random effects of model initialization, training data shuffling, and selecting the elements to omit, we conduct three runs for each
of the experiments. 
For a fair comparison, the removed items in the test set remain the same for all models during one run with a fixed pruning parameter.

We also investigate the effect of completeness of the partial input set through the number of dropped elements. 
We run multiple experiments by dropping different percentages of elements with respect to the size of the original set. We performed experiments for some given pruning thresholds on PubMed, ACM, and all subject indexing datasets, but we expect a similar behavior for other thresholds.

\section{Results}\label{sub:results}

The presentation of the results is structured along the research questions outlined in the introduction.
The first question (i),~on comparing the recommender performance in two scenarios with different underlying semantics of item co-occurrence, is jointly reflected by Section~\ref{sec:citationresults}, which covers the results from citation recommendation task, and Section~\ref{sec:labelingresults}, which presents the results from the subject label recommendation task.
Within each task, we then present the results along questions
(ii)~on how the completeness of the partial set of items influences the provided recommendations, 
(iii)~on how pruning affects the models' performance, and
(iv)~on the influence of using bibliographic metadata as input.
 
\subsection{Results for the Citation Recommendation Datasets}
\label{sec:citationresults}

\begin{figure*}[!h]
 \begin{subfigure}{\textwidth}
        \includegraphics[width=\textwidth]{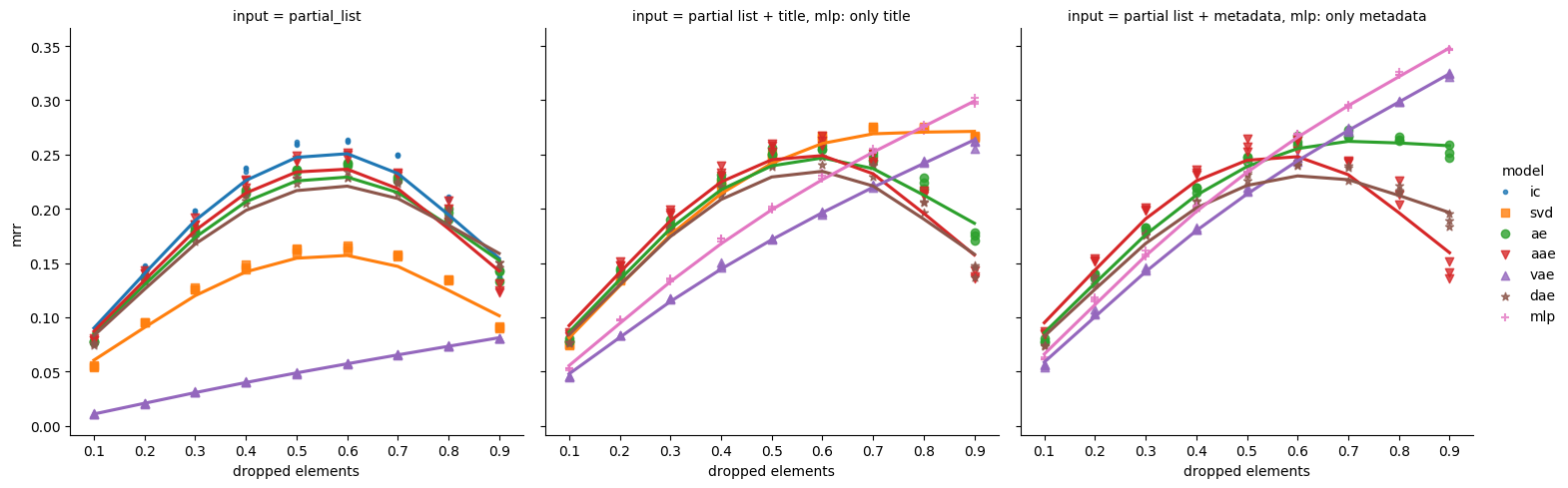}
    \end{subfigure}
    \begin{subfigure}{\textwidth}
        \includegraphics[width=\textwidth]{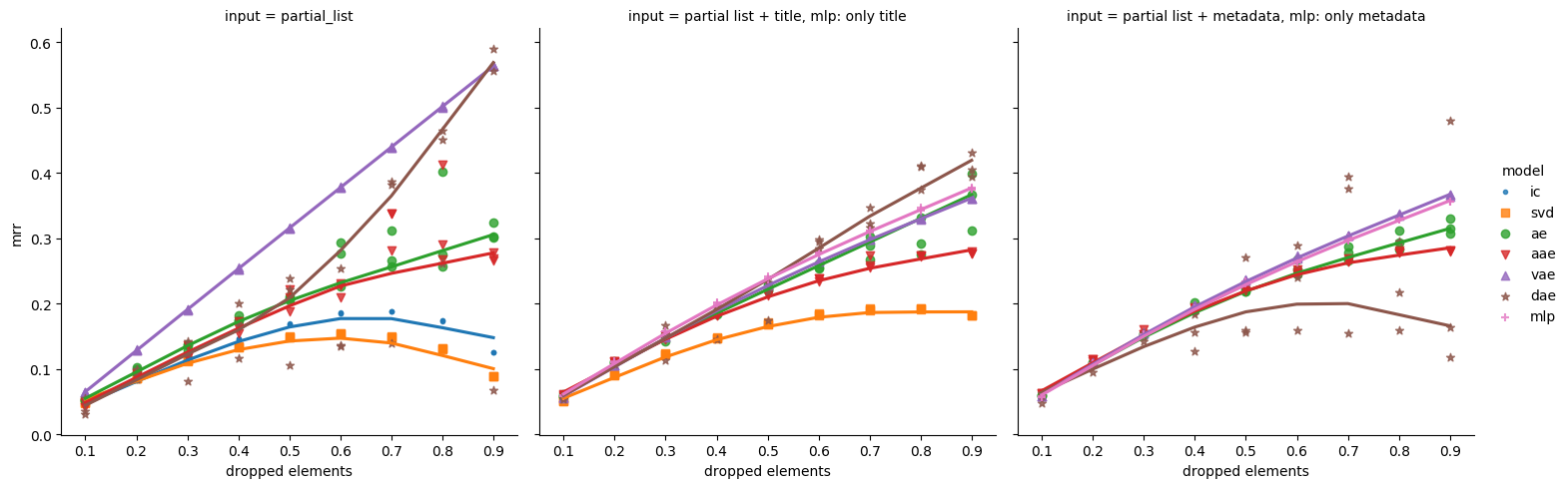}
    \end{subfigure}
    
  \caption{MRR of predicted citations on the test set with varying
  number dropped elements for the PubMed (top row) and ACM (bottom row) citation datasets. 
  The minimum item occurrence threshold is set to $55$.
  \textit{Left}: Only the partial set of items is given.
  \textit{Center}: The partial set of items along with the document title is given. 
  \textit{Right}: The partial set of items is given along with the document title, the authors, the journal title, and the MeSH labels. 
MLP can only make use of either titles or titles, authors,  journal titles, and MeSH labels.}\label{fig:pubmed:drop}
\end{figure*}

Figure~\ref{fig:pubmed:drop}
depicts the results for the models on the citation recommendation task with respect to the drop parameter that controls the percentage of dropped elements in the original sets of references, \ie{} how the completeness of the partial set of items influences the provided recommendation, on PubMed and ACM, respectively.
On PubMed, most of the models peak around a drop threshold of about 50\,\%, independently if metadata are used or not. The exceptions are MLP and VAE, which increase also with higher percentages of dropped elements. SVD with titles and AE with more metadata plateau when 60\,\% of elements are dropped.  MLP and SVD have low results with few dropped elements but achieve good performance with high drop thresholds. 
The same holds for VAE when titles or even more metadata are available.
On ACM, only item co-occurrence and SVD decrease with more than 60\,\% of elements dropped. However, different autoencoders have a lower improvement with many dropped elements, depending on their type and the additional information used (only the partial set, partial set and titles, or  partial set, titles, and more metadata).
In contrast, DAE with only the partial set tends to increase more with a drop threshold higher than 50\,\%.
The more metadata are used the lower is the improvement with many elements dropped.

\begin{figure*}[!h]
  \begin{subfigure}{\textwidth}
        \includegraphics[width=\textwidth]{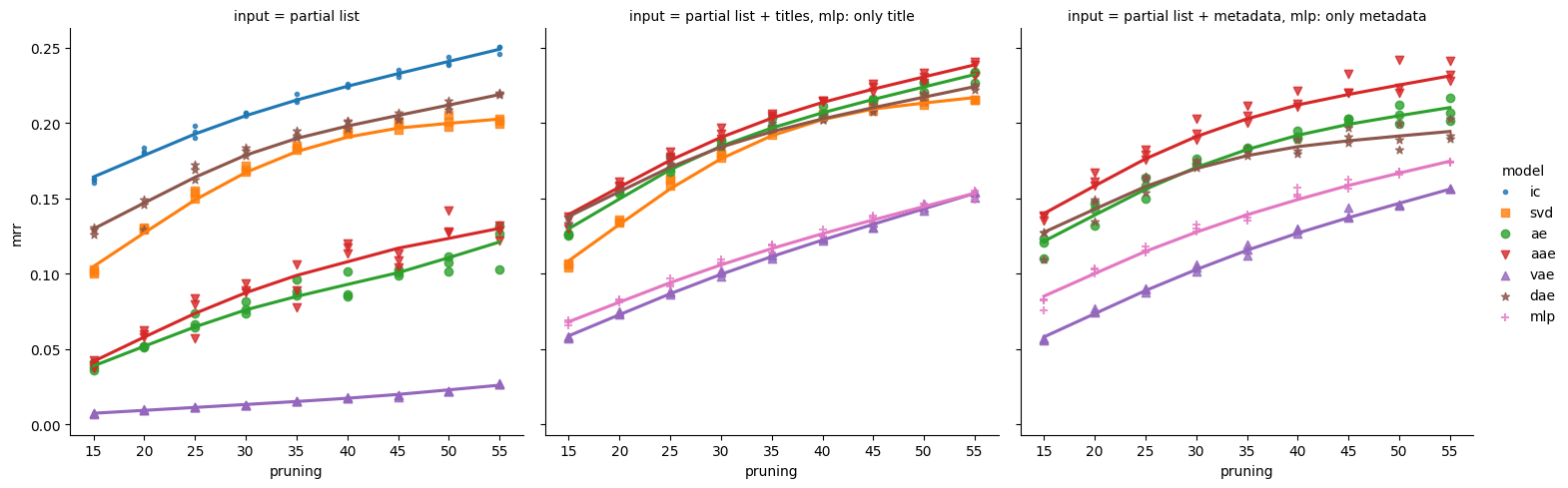}
  \end{subfigure}
  \begin{subfigure}{\textwidth}
        \includegraphics[width=\textwidth]{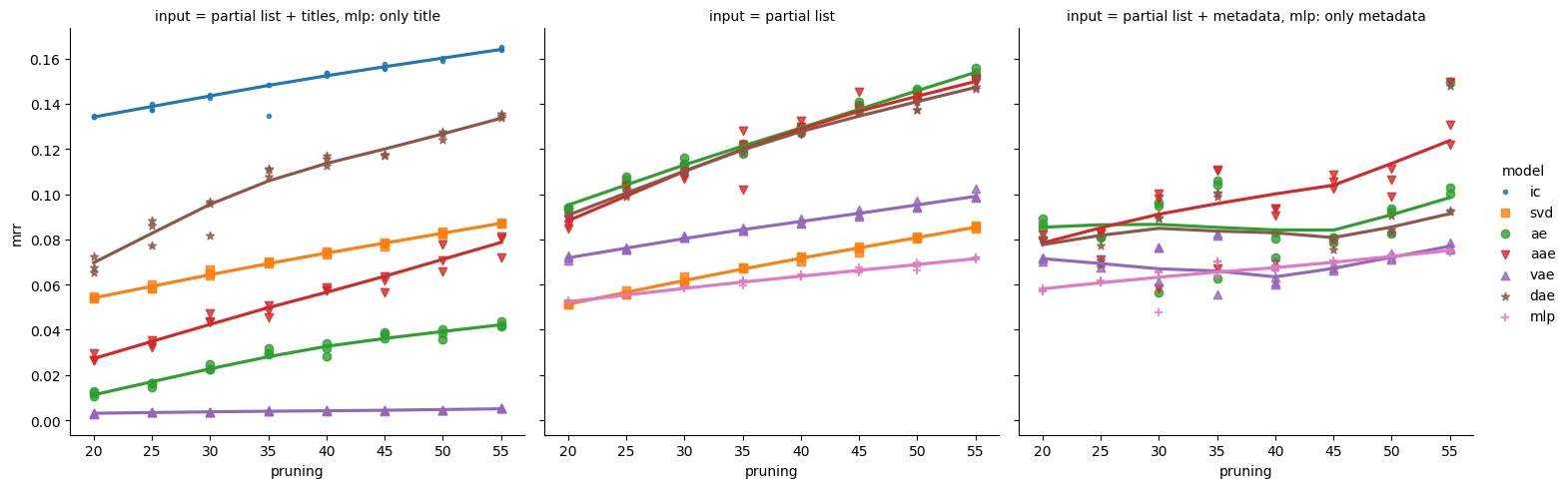}
  
  \end{subfigure}
  \begin{subfigure}{\textwidth}
        \includegraphics[width=\textwidth]{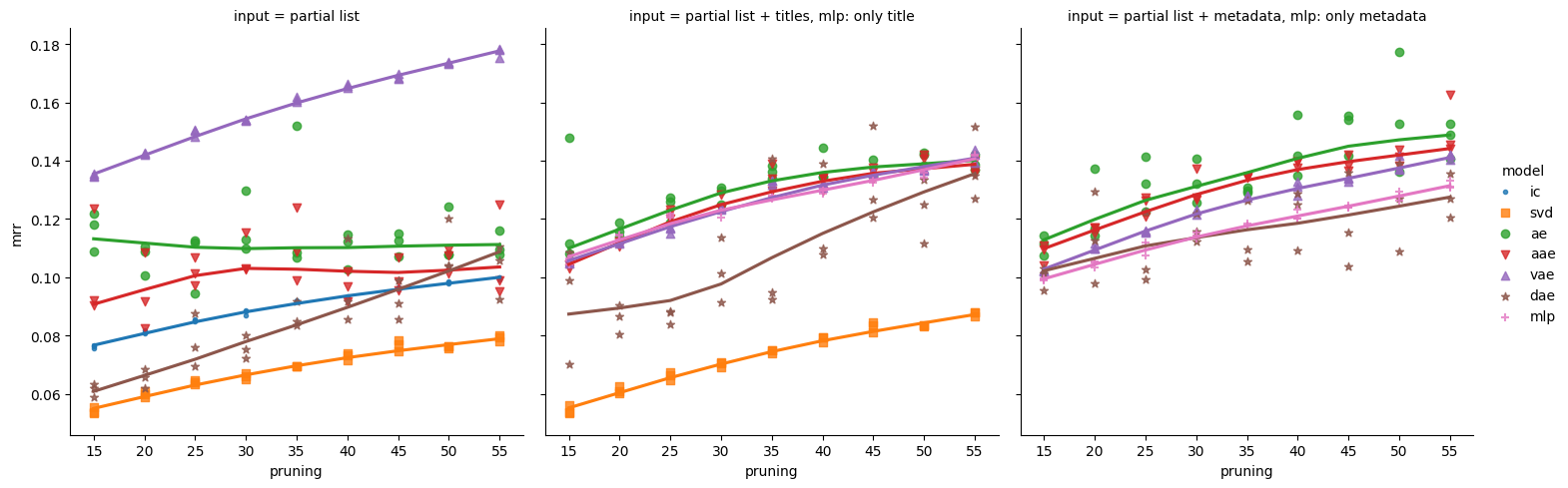}
  \end{subfigure}
  \caption{MRR of predicted citations on the test set with varying
  minimum item occurrence (pruning) thresholds for the PubMed (top row), DBLP (middle row), and ACM (bottom row) citation datasets. 
  \textit{Left}: Only the partial set of items is given.
  \textit{Center}: The partial set of items along with the document title is
  given. 
  \textit{Right}: The partial set of items is given along with the document title, the authors, the journal title and the MeSH labels (if available for a document). 
MLP can only make use of either titles or titles, authors,  journal titles, and MeSH labels.
}
 \label{fig:pubmed:results}
\end{figure*}

Figure~\ref{fig:pubmed:results}
shows the results for the models with respect to the pruning parameter that controls the number of considered items as well as
the sparsity (see Table~\ref{tab:citation:pruning}) on the PubMed, DBLP, and ACM datasets, respectively. 
We observe a trend that a more aggressive pruning threshold leads to higher scores among all models on all three datasets.
However, this phenomenon seems to be more attenuated on the ACM dataset. 
Notably, on this dataset, AE and AAE seem to be unaffected by the threshold, or even there seems to be a slight decrease for higher thresholds.
When no title information is given, the item co-occurrence approach performs best on PubMed and DBLP, while VAE obtained the best scores on ACM. 
When titles are used, autoencoders (AAE, AE, and DAE) become competitive to the item co-occurrence approach
and outperformed other models on PubMed and DBLP. The same holds when additional metadata are available.
The results on the ACM dataset show a similar pattern, except that DAE performs worse than VAE. 
Surprisingly, more metadata yield worse results than titles only.

\subsection{Results for the Subject Label Recommendation Datasets}
\label{sec:labelingresults}

\begin{figure*}[!h]
 \begin{subfigure}{\textwidth}
        \includegraphics[width=\textwidth]{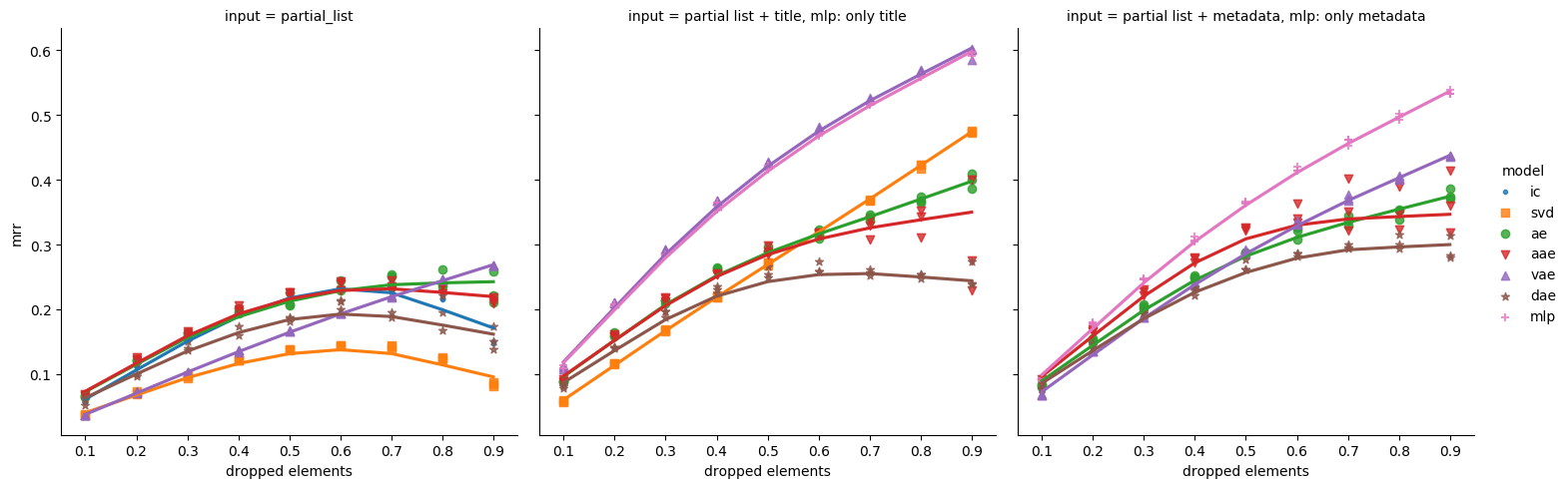}
    \end{subfigure}
    \begin{subfigure}{\textwidth}
        \includegraphics[width=\textwidth]{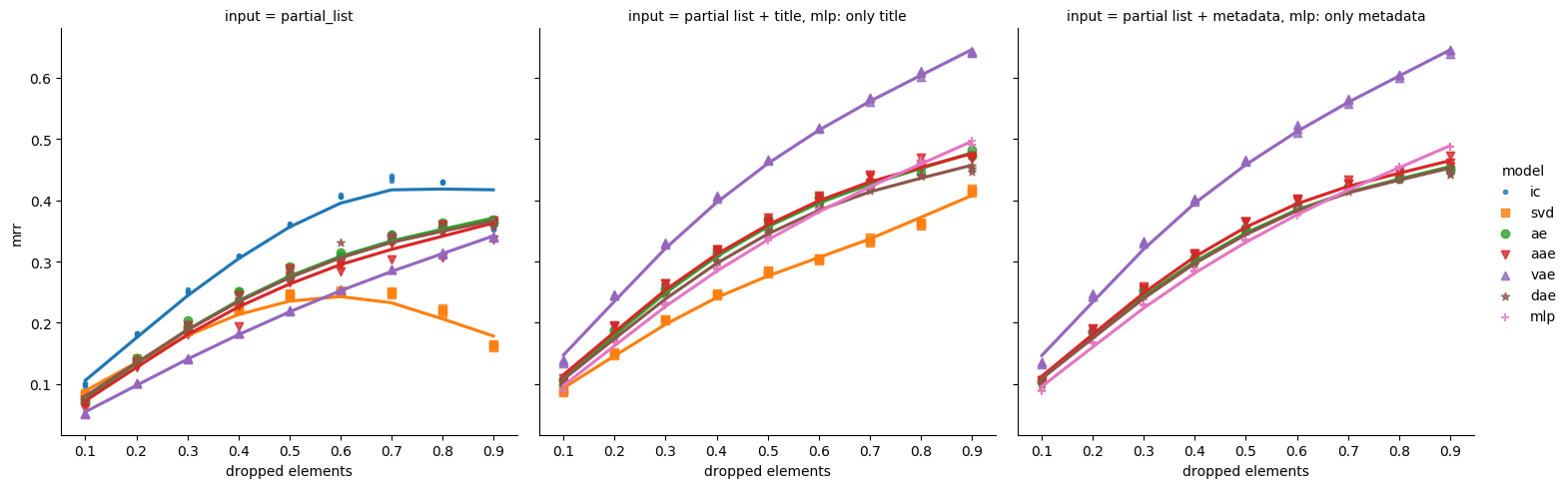}
    \end{subfigure}
    \begin{subfigure}{\textwidth}
        \includegraphics[width=0.7\textwidth]{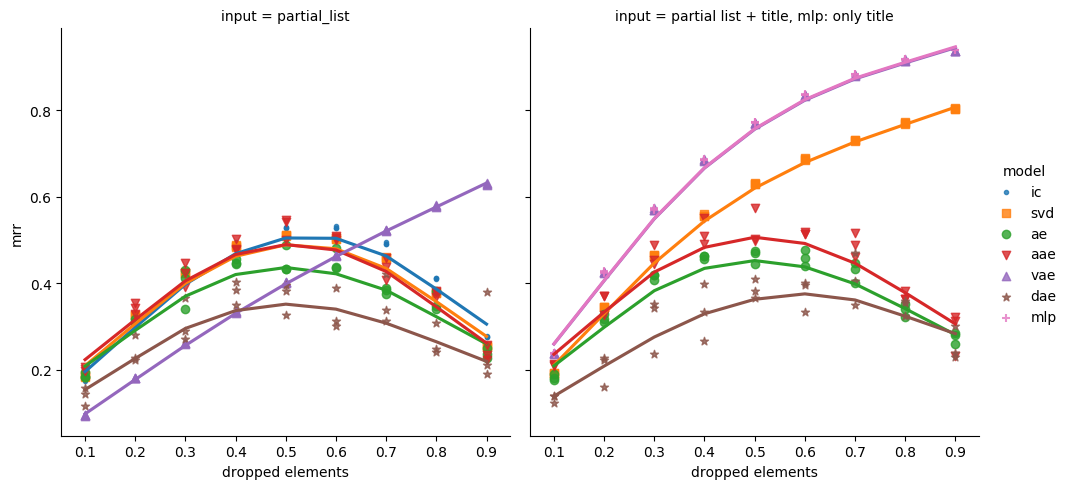}
    \end{subfigure}
    
  \caption{MRR of predicted subject labels on the test set with varying
  number of dropped elements for the EconBiz (top row), IREON (middle row) and Reuters (bottom row) datasets. 
  The minimum item occurrence threshold is set to $20$.
  \textit{Left}: Only the partial set of items is given.
  \textit{Center}: The partial set of items along with the title is given. 
  \textit{Right}: The partial set of items along with the document title and authors is given. 
  MLP can only use either titles or titles and authors.}\label{fig:economics:drop}
\end{figure*}

\begin{figure*}[!h]
  \begin{subfigure}{\textwidth}
        \includegraphics[width=\textwidth]{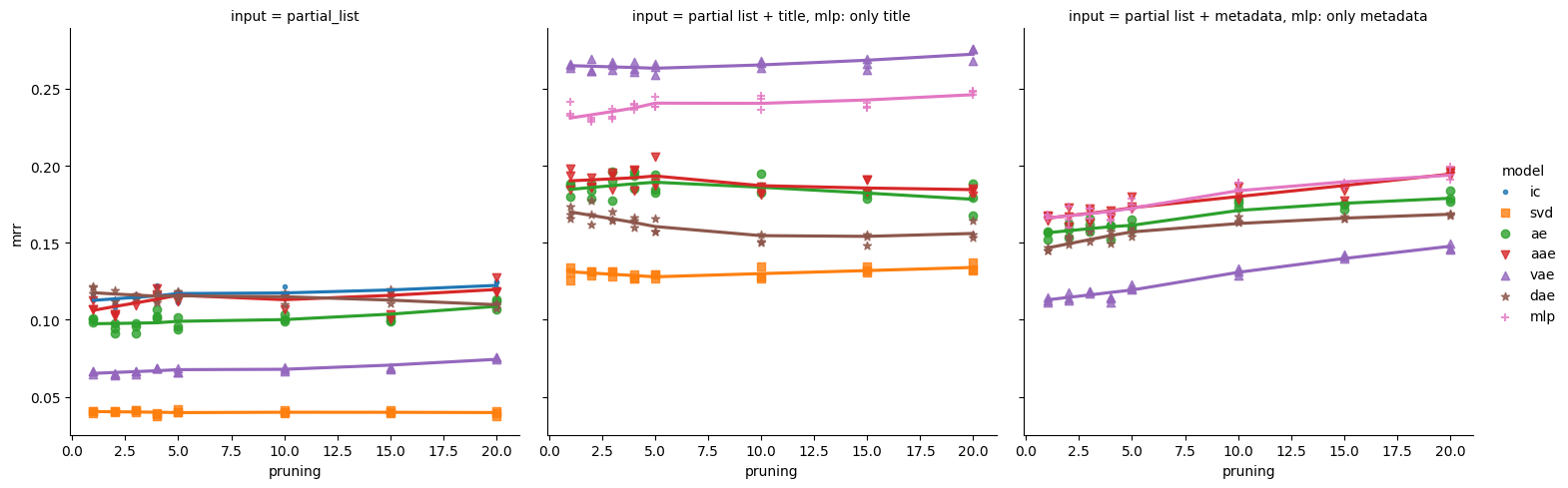}
  \end{subfigure}
  \begin{subfigure}{\textwidth}
        \includegraphics[width=\textwidth]{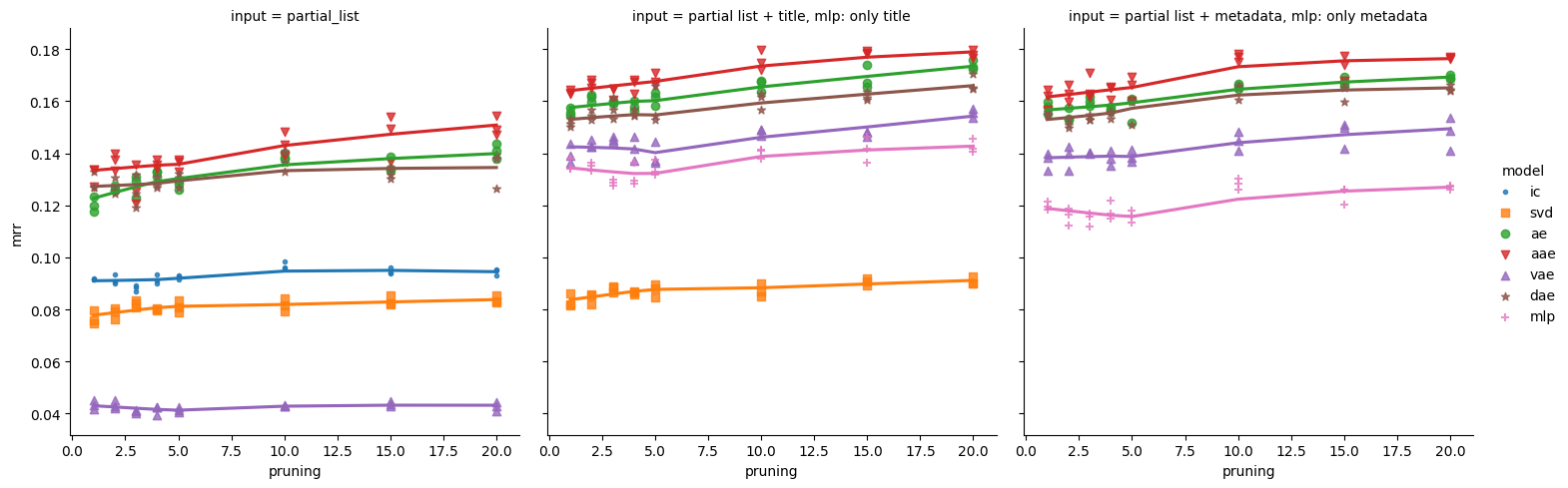}
  \end{subfigure}
  \begin{subfigure}{\textwidth}
        \includegraphics[width=0.7\textwidth]{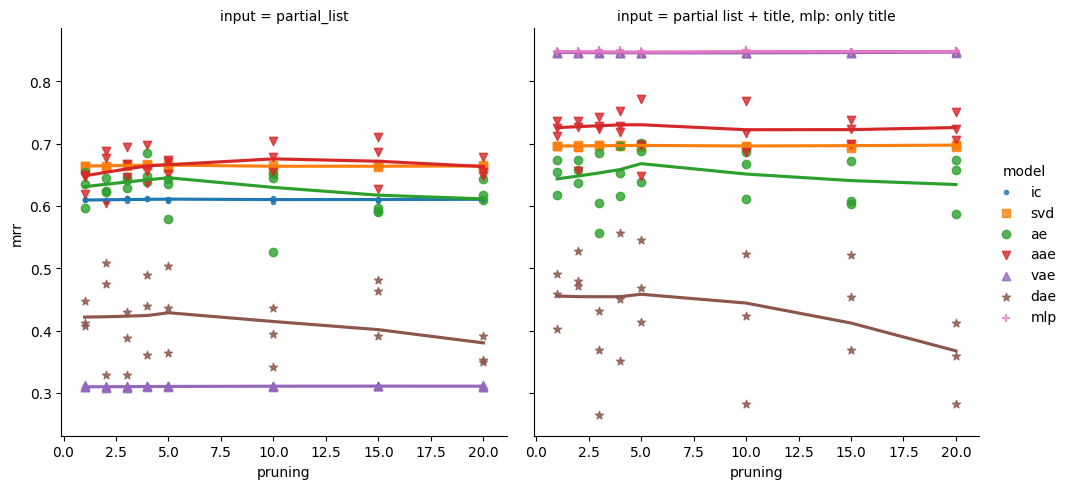}
  \end{subfigure}
  \caption{MRR of predicted subject labels on the test set with varying minimum item occurrence thresholds for the EconBiz (top row), IREON (middle row) and Reuters (bottom row) dataset. 
  \textit{Left}: Only the partial set of items is given. 
  \textit{Center}: The partial set of items along with the document title is given.
  \textit{Right}: The partial set of items along with the title and authors is given. MLP can only use either titles or titles and authors.}
  \label{fig:economics:results}
\end{figure*}

Figure~\ref{fig:economics:drop}
depicts the results for the models for the subject label task with respect to the number of dropped elements on EconBiz, IREON, and Reuters, respectively.
As for citations, we performed experiments only for some given pruning thresholds, but we expect a similar behavior for other thresholds. 
When no title information is available, most of the models peak or plateau around a drop threshold of 50\,\% and 60\,\% on Reuters and EconBiz, respectively. 
On IREON, item co-occurrence plateaus at 70\,\%, SVD peaks at 60\,\%, the other models show a steady increase.
With titles only, SVD performs poorly when few elements are dropped, while outperforms various models with many elements dropped (except for IREON).
When titles and additional metadata are given, only AE, AAE, and DAE increase considerably less with many elements dropped on EconBiz, while they peak at about 50,\% on Reuters. 
On IREON, all the models suffer less when many elements are dropped.
VAE and MLP are generally the best performing models in all datasets, although the most effective  varies with the dataset and the metadata used. 

Figure~\ref{fig:economics:results} shows the results for the models with respect to the pruning parameter that controls the number of considered items and the corresponding sparsity (see Table~\ref{tab:label:pruning}) on the EconBiz, IREON, and Reuters datasets, respectively. 
When no title information is available, autoencoders (but VAE) are competitive to the item co-occurrence approach. 
When titles are given, the models achieve considerably higher scores than all models operating without this information. 
Specifically, VAE achieves the best results, closely followed by MLP, on EconBiz, while AAE is the best-performing model on IREON.
On reuters MLP and VAE achieve the same performance. 
Similar to the citation task, using additional metadata decrease the performance compared to using only titles, although in IREON the decline is notably lower. 
VAE is particularly negatively affected by more metadata on EconBiz. 
Reuters only provides titles, so it was impossible to use additional metadata.

\section{Discussion}\label{sec:discussion}
We first recall the main findings, and then we compare the different item co-occurrence semantics in the two tasks (research question i). Subsequently, for each task, we discuss each of the remaining research questions (ii to iv).

\subsection{Main findings}
 
Regarding research question (i),
the two tasks of citation recommendation and subject label recommendation, which may seem rather similar at a first glance, are actually not.
Our experiments show that what is already cited is much more informative than which subject labels are already assigned.
This is supported by the mutual information, which is greater than $0.5$ for the citation datasets, but below or close to $0.3$ for the subject label datasets.
In the citation recommendation task, where item co-occurrence implies relatedness, the approach based on the co-occurrence is a strong baseline.
VAE (using only item sets) outperforms this baseline on ACM, but VAE performs worse when titles and other metadata are exploited.
In the subject label recommendation task, where co-occurrence of items implies diversity, item co-occurrence is not effective, and the best method depends on the dataset: AAE in IREON, VAE in EconBiz (followed by MLP), and MLP and VAE (with titles) in Reuters.
DAE seems to be less stable in optimization, as its results usually highly vary among multiple runs on the same dataset.

On research question (ii), from our experiment on the varying number of dropped elements emerged that the models which do not show the boomerang-shaped curve when using metadata are the ones that rely more on titles and other metadata than on the partial set of items.
This enables them to perform well also when very few items are given as input in both the considered tasks.

Regarding research question (iii), it is desirable to have low pruning thresholds to avoid a ``rich get richer'' phenomenon, where documents highly cited, or labels frequently assigned, are privileged. 
In fact, highly cited documents, or often used subject labels, are also the ones most likely to be known, while recommendations should also lead to discovering previously unknown items. 

Finally, about research question (iv), our results show that, counter-intuitively, it is not always better to add more metadata as input. 
Usually, titles are helpful but sometimes models which rely solely on the partial set of items are the top performers.
Titles have proved to be the most useful metadata field in our experiments.
Adding further metadata not only did not improve the performance, but sometimes even decreased them.

\subsection{Comparison of Different Semantics of Item Co-occurrence}
We observe different relationships between the factors of (i)~type of recommendation task and the performance with (ii)~varying completeness of the partial set, (iii)~degree of sparsity, as well as (iv)~input modalities. 
We discuss the results per type of recommendation task in Sections~\ref{sec:cittask} and~\ref{sec:subtask}.

Regarding the relationship research question (ii), the recommendation performance depends a lot on the completeness of the partial set. 
Most models perform best having around 50\,\% of items from the original set, but VAE and MLP achieve good performance also with 90\,\% of elements dropped. 
It is noteworthy that pruning almost never changes the order of the top performers, while the top performers differ by varying the number of dropped elements.

Second, on research question (iii), by applying several thresholds on minimum item occurrence, we controlled the number of considered items and thus the degree of sparsity. 
The result is that all considered models are similarly affected by the increased sparsity and difficulty caused by higher numbers of considered items. 
Unsurprisingly, the results decrease with lower pruning thresholds, as pruning can heavily influence the results~\cite{beel2016paper}. 
Interestingly, there are differences between the two tasks: while the decrease is pronounced in the citation datasets, it is very low in the other ones. 
This decrease is low in the subject label datasets presumably because they are considerably less sparse (see Table~\ref{tab:label:pruning} vs Table~\ref{tab:citation:pruning}). 
Differences persist even when datasets are similar. 
EconBiz has $4,568$ classes without pruning and PubMed has $4,939$ cited documents with pruning at $40$ and $3,904$ with pruning at $45$.
On PubMed (citations), the best MRR is about 21\,\% with item co-occurrence (not improving with metadata), while on  EconBiz (subject labels), all models yield similar results only when using metadata, and without metadata are below 13\,\%.

Finally, regarding research question (iv), on the citation task the partial set of citations is the most important information to recommend potentially missing citations. 
For the subject label recommendation task, however, the MLP model, which exploits titles only, achieves the best performance in one dataset and is generally competitive.
On the citation recommendation task, autoencoders and SVD  become competitive to the strong  co-citation baseline when titles.

\subsection{Detailed Discussion of the Citation Recommendation Task}\label{sec:cittask}

The number of elements dropped in the partial input sets differently affects the models.
Most of the models improve until 
about
50\,\% of elements are dropped, then their performance decreases.
In fact, the more elements are dropped less information is used as input, but the task becomes also easier as there is more than one correct answer (one document or label to predict). 
When metadata are provided together with the partial set, the results generally improve. 
This is particularly the case when the partial input set is small, i.\,e., many elements are removed, as additional information compensates for fewer items in the partial set. 
MLP results improve also with many elements dropped as it does not exploit the partial set of items. 
VAE shows similar behavior. 
It is capable of providing good predictions even when few items are given as input. 
This could be because of the generalization provided through the Gaussian prior to the code.
The reason is that the latent representations learned by distinguishing the code from a smooth prior make the model more robust to sparse input vectors as smoothness is key for good representations that disentangle the explanatory factors of variation~\cite{bengio2013representation}.
Although both VAE and AAE impose a Gaussian prior on the code, VAE is better in some cases. 
A possible explanation is that VAE is more stable during optimization~\cite{TolstikhinBGS18}.
Lastly, on ACM the results continue to improve even with very few items in the partial set (i.\,e., many elements are dropped), except for item co-occurrence and SVD.
This may be due to ACM characteristics. 
First, the mutual information for ACM ($0.5282$) is the lowest among the citation datasets ($0.5407$ for DBLP and $0.5996$ for PubMed).
This means that already cited documents are less informative in ACM compared to the other two datasets.
Second, the difference among the number of most cited and less cited documents is lower in ACM ($700$) than in the other two datasets ($5,000$ in DBLP and $16,000$ in PubMed), as shown by the y-axes in Figure~\ref{fig:docs_stats}. 
So more documents have many references.
In this case, although many references may have been dropped, more documents still may have enough references for the model to learn something.

Regarding the results for the experiments with varying pruning, co-citation is known to be an important baseline~\cite{DBLP:journals/jasis/Small73}.
However, we have shown that is not always the case.
For the ACM dataset, VAE is the best method.
This is interesting, because in other cases VAE particularly underperforms on DBLP and PubMed, especially without metadata. 
This may be due to ACM special characteristics, as already discussed for the number of dropped elements. 
Specifically, already cited documents are less informative in ACM compared to the other two datasets due to the lower mutual information.

On the influence of metadata, MLP and autoencoders have the advantage of exploiting metadata information (autoencoders along with the partial set of citations), which improves the results.
The use of titles is most effective.
The use of other metadata attributes, such as authors and venues, on top of titles, does not improve the recommendation performance.
%
From the partial input set, autoencoders can learn co-occurrence within item sets and may also learn to put appropriate weights in the bias parameters if it is helpful for the overall objective. 
In this task of citation recommendation, the co-occurrence of documents within item sets is of great importance as related papers are typically cited together, which explains the strength of the item co-occurrence baseline.
Although MLP can also learn bias in the output layer to model the items' prior distribution, autoencoders can model the relation between cited documents from the partial input set.
MLP cannot model the relation between cited documents because it only uses the titles as input and not the partial set.
When considering different types of metadata in the case of MLP, using titles is generally as effective as, or even more effective than, using additional metadata available (together with the titles).
We think this is because title is more indicative than other metadata when deciding whether to cite a paper.
%
Although researchers may tend to cite more often some papers from well-known venues as well as some authors, because of similarity in topics or because they know them, the title has a stronger relation to the paper than the authors and venues.
Furthermore, the advantage of the title is that this metadata is always available, even in the news domain, while other metadata are not. 
For example, 
in PubMed only about 77\,\% of the documents have MeSH labels, in ACM roughly 93\,\% of documents include authors, 
and in DBLP around 83\,\% of papers hold the venue (see Table \ref{tab:metadata}), while for all documents there is a title.
 
\subsection{Detailed Discussion of Subject Label Recommendation Task}\label{sec:subtask}

The number of elements dropped has a lower effect on the models compared to the citation recommendation task.
Most of the models improve until when about 50\,\% of the elements in the original set are dropped, then their performance plateaus or slightly decreases.
Nevertheless, some models can provide good predictions also with few elements in the partial set.
As in the citation recommendation task, results of MLP and VAE improve even with many elements dropped.
Only with the Reuters dataset, many models show a notable decline.
This could be due to its distribution of the label occurrence, which is the only one that does not follow the power-law distribution.
On the contrary, there is a low number (roughly $100$ compared to $4,600$ in EconBiz and $10,000$ in IREON) of fairly well-balanced labels to choose from (see Figure \ref{fig:label_stats}).
Another reason could be that in the Reuters dataset the labels have no hierarchy, contrary to the other two subject label datasets that are based on a professional taxonomy.
Subject indexers usually assign the ancestor instead of the child subjects when two or more subject labels with a common ancestor in the hierarchical thesaurus match~\cite{DBLP:conf/kcap/Grosse-BoltingN15}.
Thus, two subjects that are semantically related because they share a common ancestor are unlikely to co-occur in the annotations of a single document. 
Without a hierarchy, different subject labels can be similar and no common ancestor is available to use instead.
For example, the news ``\emph{Clinton signs law raising minimum wage to \$5.15}'' is assigned the subjects \emph{EMPLOYMENT/LABOUR}, \emph{LABOUR}, \emph{LABOUR ISSUES} which are rather similar.
Finally, as the main purpose of subject indexing is retrieval, the fact that there are many labels used, often suggests that recall is preferred over precision.
While in a library it may be desirable to retrieve fewer results but all highly relevant to a query,
in the news domain it may be best to retrieve as many results as possible, although some could be less relevant.

In the experiments with varying pruning, MLP and VAE achieve the highest mean reciprocal rank in two out of three datasets.  
Thus, already assigned subjects are less informative for a subject label recommendation task than the titles are.
Two subjects that are semantically related because they share a common
ancestor is unlikely to co-occur in the annotations of a single document since subject indexers tend to rely on the ancestor instead of the child subjects.
On the IREON dataset, instead, autoencoders and notably AAE, are most effective. 
A reason might be that IREON has much more different subject labels than the other datasets (see Table \ref{tab:label:pruning}). 
Therefore, co-occurrence statistics (which are modeled by AE variants) might be more informative.

Regarding the influence of metadata, adding other metadata with titles generally decreases the performance. 
As for citations, other metadata may be less indicative of the paper content. 
Furthermore, only the authors are available in addition to titles. 
Authors are present in roughly 98\,\% and 83\,\% of documents, in EconBiz and IREON, respectively.
VAE is the best model (together with MLP) in Reuters, but performs poorly without titles. 
This suggests that when provided with titles, VAE learns to ignore the input item set, which is not possible with only the partial set of items as input.

\subsection{Threats to Validity}
The availability and occurrences of metadata fields varies among the datasets, as shown in Table~\ref{tab:metadata}. 
Citations datasets have more metadata attributes available than the subject label datasets. 
Titles and authors can be used in all three citation datasets; in PubMed also journal and MeSH labels can be exploited, and venues are present in ACM and DBLP. 
For the subject label datasets, only titles and authors can be used, apart from Reuters which offers only titles. 
So it was only partially possible to use the same or similar set of metadata fields among the datasets. 
Notably, we could not use abstract information because it was always or often missing. 
Only ACM and DBLP contain such information but it is rarely available in ACM (less than 4\,\%), so we decided not to use it in our experiments.

We tested the models' performance and the impact of different semantics of item cooccurrence, the completeness of partial input set, the pruning, and metadata on various datasets, for each of the two recommendation tasks. 
As the results are consistent among the datasets, we have good reasons to assume that the results can also be generalized to other, similar datasets in the considered domains. 
Our method to add metadata is general, and can handle different types of metadata. 
Thus, the models can also be applied to similar tasks in other domains, which could also benefit of the use of metadata. 
For example, we have previously shown that metadata are beneficial for automatic playlist continuation~\cite{DBLP:conf/recsys/VaglianoGMS18}. 

\subsection{Practical Implications}

Our experiments have a high practical relevance as they are close to real-world settings. 
This comes in three aspects.
First, we use six real-world datasets from five different domains.
Thus, the experimental results show how the recommender models would behave in real-world contexts.
Second, the splitting of the documents in training and test set along the time axis resembles the natural constraint that newly written publications can only cite already published works and only paper published before are already annotated. Additonally, applying this chronological split to subject labels also account for concept drift~\cite{DBLP:journals/datamine/WebbLGP18}.
Third, by taking into account the typical long-tail distribution in users' feedback to items (in our case, citations and label annotations), we have also investigated performance with low pruning thresholds, i.\,e., including items with few citations and labels rarely used to annotate documents.
This further strengthens the reliability of our experimental results in real-world settings, in contrast to existing studies with limited datasets induced by pruning items based on frequency.

\section{Conclusion}\label{sec:conclusion}

Different semantic interpretation of item co-occurrence in recommendation tasks highly affects the preferable input modalities. 
When item co-occurrence resembles relatedness, such as in citations, the set of already cited documents is beneficial.
In subject recommendations, co-occurring subject labels do not imply that these subjects are similar. 
Instead, a document's actual research subject needs to be described by using multiple, diverse subject labels as annotations. 
Incorporating multiple input modalities offers a conceptual benefit, but not always adding more metadata is useful, and relying on just titles or the partial set of items can be more effective.
All of the evaluated methods are similarly sensitive to data sparsity, but variational autoencoders and multi-layer perceptron are more robust with few items in the partial set.
This is likely because they rely more on titles and other metadata than on the initial item set.

In future systems and studies, the semantics of item co-occurrence should be taken into account when deciding whether the partial list of items should be supplied to a recommendation model as input.
Regarding the two scenarios considered, we can state:
In citation recommendation, where co-occurring items are similar, model can perform well without using additional metadata. 
In subject indexing, where co-occurring items are diverse, using the content is more effective than using the partial item set.
Investigating further recommendation scenarios with respect to the influence of item co-occurence semantics to the recommendation performance remains part of future work.

\section{Declarations}

\textbf{Funding}
This work was partially supported by the EU H2020 project MOVING (contract no 693092).\\
\textbf{Conflicts of interest}
We have no competing interest to report.\\
\textbf{Availability of data and material}
Not applicable\\
\textbf{Code availability}
The source code for reproducing and extending our experiments is openly available on GitHub: \url{https://github.com/lgalke/aae-recommender}.\\
\textbf{Acknowledgement}
We thank Gunnar Gerstenkorn for his support with the data preparation and preprocessing.

\bibliographystyle{elsarticle-num} 
\bibliography{references.bib}

\end{document}